\documentclass[sigconf]{acmart}
\AtBeginDocument{%
  \providecommand\BibTeX{{%
    \normalfont B\kern-0.5em{\scshape i\kern-0.25em b}\kern-0.8em\TeX}}}

\usepackage{xcolor}

\newcommand{\change}[1]{{\textcolor{black}{#1}}}

\usepackage{xspace}
\newcommand{\system}{Script\&Shift\xspace} 
\newcommand{\systems}{Script\&Shift's\xspace}

\usepackage{fontawesome}
\usepackage{tikz}
\usepackage{graphicx}
\usepackage{subcaption}

\definecolor{skyblue}{RGB}{135, 206, 235}
\definecolor{lightestgray}{RGB}{230, 230, 230}
\definecolor{legalyellow}{RGB}{255, 255, 204}

\definecolor{ToneTara}{RGB}{52, 130, 206} 
\definecolor{IdeaIvy}{RGB}{128, 0, 128}  
\definecolor{DetailDanny}{RGB}{249, 186, 97} 
\definecolor{FeedbackFelix}{RGB}{34, 211, 238} 
\definecolor{AudienceAli}{RGB}{255, 102, 26} 
\definecolor{StructureSam}{RGB}{123, 183, 116} 

\newcommand{\coloredcircle}[1]{%
  \tikz\draw[fill=#1, draw=none] (0,0) circle (1.2ex); 
}

\newcommand{\pageiconwithnumber}[3][black]{%
  \raisebox{-.1\height}{
    \begin{tikzpicture}
      \node[inner sep=0pt, text=#1] {\faFile}; 
      \node[align=center, inner sep=0pt, text=#2] at (0,0) {\normalsize\textbf{#3}}; 
    \end{tikzpicture}%
  }%
}

\newcommand{\customdashedbordertext}[3]{%
  \tikz[baseline=(X.base)] \node[draw=#2, rectangle, dashed, fill=#3, fill opacity=0.1, text opacity=1, inner sep=2pt, rounded corners] (X) {\color{black}#1};%
}

\newcommand{\customsolidbordertext}[3]{%
  \tikz[baseline=(X.base)] \node[draw=#2, rectangle, fill=#3, fill opacity=0.1, text opacity=1, inner sep=2pt, rounded corners] (X) {\color{black}#1};%
}

\newcommand{\twopageicon}[2]{%
  \begin{tikzpicture}[baseline=(current bounding box.south)]
    \draw[fill=#1, draw=black] (0, 0) rectangle ++(1em, 1.2ex);
    \draw[fill=#2, draw=black] (0, -1.5ex) rectangle ++(1em, 1.2ex);
  \end{tikzpicture}%
}

\copyrightyear{2025}
\acmYear{2025}
\acmConference[CHI '25]{CHI Conference on Human Factors in Computing
Systems}{April 26-May 1, 2025}{Yokohama, Japan}
\acmBooktitle{CHI Conference on Human Factors in Computing Systems (CHI
'25), April 26-May 1, 2025, Yokohama,
Japan}\acmDOI{10.1145/3706598.3714119}
\acmISBN{979-8-4007-1394-1/25/04}

\begin{document}

\title[A Layered Interface for Writing with LLMs]{\system: A Layered Interface Paradigm for Integrating Content Development and Rhetorical Strategy with LLM Writing Assistants}


\author{Momin Siddiqui}
 \affiliation{%
  \institution{Georgia Institute of Technology}
   \country{USA}
 }
 \email{msiddiqui66@gatech.edu}

\author{Roy Pea}
 \affiliation{%
  \institution{Stanford University}
   \country{USA}
 }
 \email{roypea@stanford.edu}

\author{Hari Subramonyam}
 \affiliation{%
  \institution{Stanford University}
   \country{USA}
 }
 \email{harihars@stanford.edu}

\renewcommand{\shortauthors}{Siddiqui, et al.}

\begin{abstract}
  Good writing is a dynamic process of knowledge transformation, where writers refine and evolve ideas through planning, translating, and reviewing. Generative AI-powered writing tools can enhance this process but may also disrupt the natural flow of writing, such as when using LLMs for complex tasks like restructuring content across different sections or creating smooth transitions. We introduce \system, a \textit{layered interface paradigm} designed to minimize these disruptions by aligning writing intents with LLM capabilities to support diverse content development and rhetorical strategies. By bridging envisioning, semantic, and articulatory distances, \systems interactions allow writers to leverage LLMs for various content development tasks (\textit{scripting}) and experiment with diverse organization strategies while tailoring their writing for different audiences (\textit{shifting}). This approach preserves creative control while encouraging divergent and iterative writing. Our evaluation shows that \system enables writers to creatively and efficiently incorporate LLMs while preserving a natural flow of composition.
\end{abstract}


\begin{teaserfigure}
  \includegraphics[width=\textwidth]{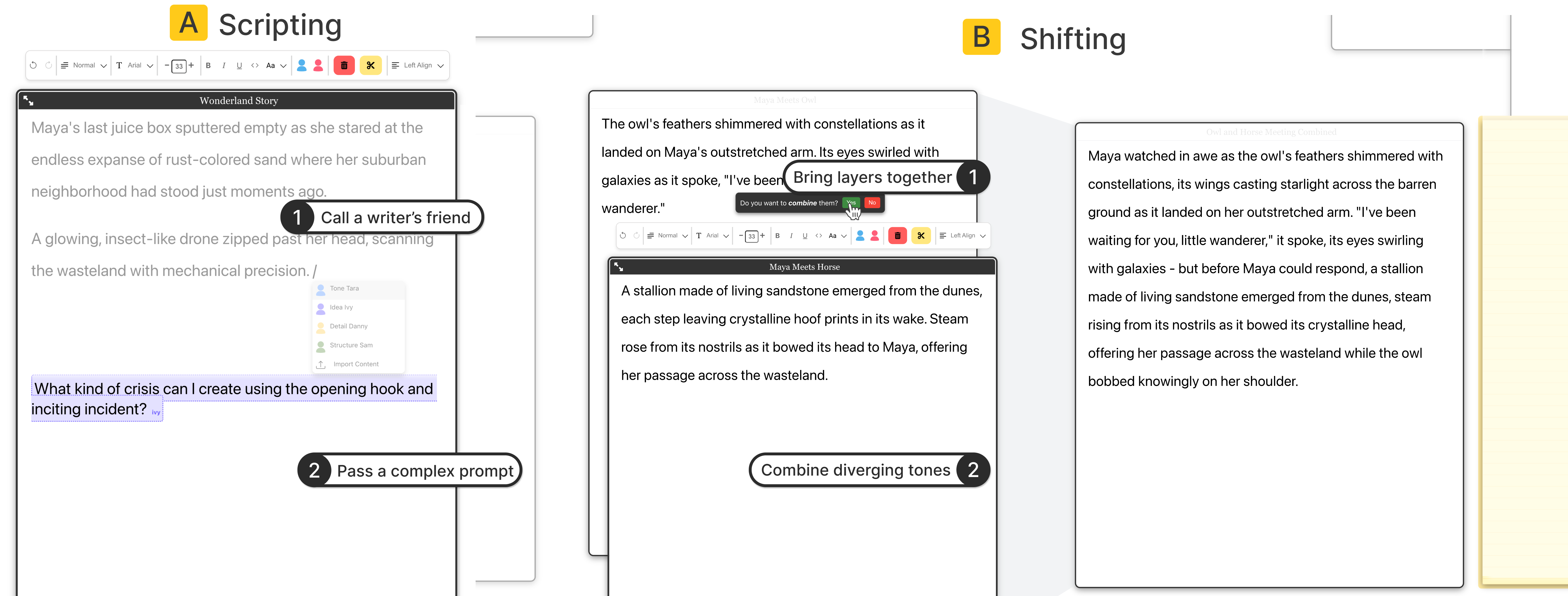}
  \caption{\system is an AI-assisted writing interface that empowers writers to query LLMs using familiar visual design elements. (A) Scripting enables users to query several specialized Writer's Friends inline. (B) Shifting allows writers to reorganize layers and combine them to see what the final document looks like. }
  \Description{}
  \label{fig:teaser}
\end{teaserfigure}


\maketitle

\section{Introduction}
Writing is rarely a linear process. Instead, it involves a complex interplay of cognitive activities  spanning various levels of abstractions and perspectives~\cite{kellogg1999psychology, pea1987chapter}. Writers must juggle fine-grained details --- such as word choice and sentence structure --- while simultaneously managing broader concerns such as overall organization and rhetorical strategy~\cite{hayes2012modeling, hayes1980identifying}. For instance, imagine writing the introduction to this CHI paper. The author might start by crafting a compelling opening sentence (e.g., \textit{``Large Language Models (LLMs) are effective writing collaborators.''}), only to pause and reconsider how this framing aligns with the paper's overall contribution. They might zoom out to then enumerate  what the key contributions are before diving back in to flesh out a particular argument. As they write, they must anticipate potential questions and critiques from readers, adjusting their narrative to address each of these imagined interlocutors. This dynamic interplay between knowledge creation and rhetorical problem-solving is what distinguishes good writing as a process of \textit{knowledge transformation} rather than mere knowledge telling~\cite{bereiter1987psychology}. 

However, the quality of \textit{writing tools} can make or break this dynamic process of composition, largely depending on the affordances they provide -- or fail to provide~\cite{kozma1991computer, kellogg1999psychology}. Tools such as pen and paper afford a flexible tactile medium for writing, but quickly become limiting when writers need to make extensive revisions or rearrange large blocks of text. Word processors, such as Microsoft Word, allow easy editing and formatting, but the page-based metaphor makes it challenging to escape the rigid top-to-bottom organization. For instance, in the CHI paper example, the author may want to try multiple ways of presenting the argument, such as structuring it around the design process, or components of a theoretical framework. Similarly, they may want to explore alternative arguments to benefits of LLMs in writing. However, such a process can be labor intensive and require duplicating sections, manually copying content, or creating entirely new documents just to assess how these alternative structures and arguments might develop. As a third alternative, platforms such as Overleaf allow writers to manage large sections of papers more efficiently through modular organization and referential code-like syntax. On the other hand, the precise formatting requirements can hinder exploratory writing. Ultimately, writers are left with tools that offer affordances for certain tasks, but which hinder accomplishing others. As a result, writers often avoid engaging in iterative, exploratory, and divergent writing, even though exploratory writing \cite{elbow1998writing,bean2021engaging} can produce good writing~\cite{bereiter1987psychology}. 

As evidenced by rapidly expanding landscape of LLM writing tools~\cite{Lee2024design}, LLMs hold the potential to alleviate many of the process-related challenges, but interface issues largely remain. At the process level, writers can prompt LLMs to generate alternative phrasings, improve sentence clarity, or provide stylistic variations, enabling more efficient iteration on the small details without becoming bogged down in manual rewriting~\cite{reza2023abscribe}. For instance, while drafting a section of this CHI paper, the author could request the model to suggest multiple ways to introduce a complex argument, allowing for immediate exploration of tone and clarity. Or they might use LLMs to receive feedback from different perspectives~\cite{benharrak2024writer} as well as structure suggestions for their text.

Despite these benefits, the exceptional potential of LLMs is often \textit{constrained} by the traditional, linear paradigms of existing writing tools or multi-turn conversational interfaces. These LLM environments don't naturally support the fluidity required for expansive writing. For instance, prompting LLMs across content levels remains a challenge due to these rigid document structures. A writer may wish to ask an LLM to refine a sentence within a section while simultaneously considering how that change affects the cohesion in a different section of the paper. To achieve this aim, they would need to prompt the LLM twice: first, referencing the specific sentence or paragraph for fine-grained revision, and then referencing the broader section or entire document to assess how the changes influence overall coherence. This process can be challenging because current writing tools often do not support seamless cross-referencing between micro and macro levels. Writers must manually manage the integration of these prompts, which disrupts the flow ballistics of the writing process. Therefore, the motivating question for our work is: \textit{How can writing tools be designed to better support seamless, iterative prompting across both micro and macro levels without disrupting the writer's workflow and associated cognitive flow? }

We introduce \system, a novel layered interface paradigm for LLM-augmented writing. \system is designed to address the limitations of traditional writing tools by offering a layered, zoomable workspace that allows writers to seamlessly interleave content development and structural organization. Our layered paradigm supports flexible prompting across both micro and macro levels, allowing users to issue complex queries to LLMs for content generation, restructuring, and refinement. As shown in Figure~\ref{fig:teaser} \system reduces the gulf of envisioning~\cite{subramonyam2024bridging} and execution~\cite{hutchins1985direct} by offering intuitive interactions such as inline friends, contextual toolbars, intelligent structuring, and other layer-based user interactions such as stacking and folding. These features make it easier for writers to issue complex prompts to LLMs and receive targeted feedback, streamlining the management of complex writing tasks. Results from our usability study and comparative evaluation show that \system supports expressive and exploratory writing with LLM assistants. 

Our main contributions include:
\begin{enumerate}
    \item A \textbf{Layered Interface Paradigm} for co-writing with LLMs, supporting flexible transitions between content creation and structural organization.
    \item A \textbf{Generalizable Architecture for Layered Interaction}, enabling dynamic and intuitive LLM prompting across different document based tasks.
    \item A comprehensive \textbf{Evaluation} demonstrating the system's effectiveness in enhancing writing workflows compared to chat- and page-based paradigms.
\end{enumerate}

\section{Related Work}

\subsection{Process of Writing}
Writing is a non-linear process involving three key phases: planning, translating, and reviewing~\cite{hayes1980identifying}, with writers often moving back and forth between these processes\change{~\cite{flower1980cognition}}. Observational studies have shown that writers generate high-level goals and supporting sub-goals as they develop their sense of purpose, and often adjust these goals based on new insights gained during the act of writing\change{~\cite{flower1980cognition}}. These cognitive process observations reflect the idea that writing, much like other creative design tasks such as art or music, involves learning through a reflective conversation with the materials being produced~\cite{bamberger1983learning}.

Revision is a critical, recursive process in writing that benefits from supportive interfaces~\cite{fitzgerald1987research, seow2002writing}. Advanced writing often transitions from the basic knowledge-telling model, which relies on cues from topics and discourse schemas, to a more complex, knowledge-transforming model~\cite{bereiter1987psychology}, where coherence and rhetorical development become central. The act of writing requires continuously updating one's rhetorical situation, including considerations of audience and purpose~\cite{flower1980cognition}. This insight emphasizes that having strong ideas does not automatically result in well-written prose\change{~\cite{flower1980cognition}}, necessitating tools that assist writers in refining both content and structure.

Existing writing interfaces, typically designed around a single text editor layout, have been extensively studied~\cite{whiteside1982people, egan1982learner}. However, general-purpose word processors often fail to address the specific needs of professionals and creatives, who require more robust tools for maintaining consistency, managing dependencies, and organizing related information in structured documents~\cite{han2020designing}. These tools force users to rely heavily on memory to manage tasks crucial to advanced writing, such as tracking coherence across sections and adapting content to evolving goals~\cite{han2020textlets, han2020designing}. Our work builds upon these insights, addressing the gaps in current writing tools by introducing \system, a layered interface paradigm designed to support non-linear, iterative writing processes.

\subsection{LLM Co-Writing Interfaces}
In exploring LLMs for writing~\cite{brown2020language, kenton2019bert, min2023recent, radford2018improving}, substantial work has demonstrated various ways in which LLM co-writing interfaces can be designed~\cite{Lee2024design}. While chat-based LLM interfaces have become ubiquitous and accessible~\cite{google2023gemini, openai2022chatgpt, anthropicclaude}, there remains a significant gap in effectively prompting these systems for usable responses and guiding the iterative process of generating meaningful textual outputs~\cite{zamfirescu2023johnny, kim2023towards}. Current LLM systems often lack the nuanced supports needed for writers to steer the process and maintain control over their creative work.

Several LLM co-writing systems have emerged to bridge this gap~\cite{biermann2022tool, buschek2021impact, chung2022talebrush, dang2023choice, gero2022sparks, gero2023social, giray2023prompt, goodman2022lampost, kim2023towards, lee2022coauthor, longtweetorial, mirowski2023co, shakeri2021saga, singh2023hide}. \change{While some tools focus on specific aspects of the writing process, such as brainstorming~\cite{petridis2023anglekindling, calderwoodQGC20}, content transformation~\cite{du2022read, Arnold2021GenerativeMC}, and feedback~\cite{weber2024legalwriter, hui2023lettersmith}, others offer broader support across multiple writing tasks~\cite{reza2023abscribe, kim2024diarymate, gero2022sparks, Arnold2021GenerativeMC}.}

Despite their utility, there are persistent concerns about content ownership~\cite{hoque2024hallmark}, and studies suggest that users who rely less on LLMs tend to produce writing with higher lexical density and coherence~\cite{shibani2023visual}. Furthermore, users often desire greater control over the narrative and the AI's output~\cite{ippolito2022creative, poddar2023ai}, as LLM tools have been shown to influence the writer’s opinions, raising concerns about the extent to which language models shape user thinking~\cite{jakesch2023co}. Additionally, while LLMs excel at generating content, they often fall short in creative tasks, which is why systems should focus on preserving the user's creative control, as seen in our work with \system~\cite{chakrabarty2024art}.

Recent discussions have also explored the implications of students using LLMs for academic writing, which may hinder the development of critical writing skills~\cite{bowman2022new, meyer2023chatgpt, flower1980cognition}. There is a growing consensus that the human writer should lead the process, with the LLM handling more mundane tasks~\cite{wan2022user}. This perspective highlights the importance of designing systems where AI augments, rather than dominates, the writing process.

There are also opportunities to explore tools that provide constrained support for late-stage brainstorming, planning, and reviewing, particularly for specific writing genres such as storytelling~\cite{gero2022design}. Additionally, designing AI systems that help writers understand the model’s capabilities, adapt to individual writing styles, and offer authentic feedback will be vital~\cite{gero2023social}. \change{There is a growing trend toward collages for the design of AI writing tools. \citet{buschek2024collage} identifies four key dimensions of collage-based designs: fragmentation of content, juxtaposed voices, multiple sources, and shifting roles. In their analysis, VISAR ~\cite{zhang2023visar} supports the Collage form factor best (Collage Factor=10). VISAR propose a system for argumentative writing that combines visual programming and rapid prototyping. While VISAR implements several collage dimensions, its single-page writing interface limits quick comparisons of variations. The design framework in \citet{kim2023cells} also uses text fragmentation where each separate fragment can transform or generate text. Also, there are more spatial interfaces for writing~\cite{lin2024rambler, lu2018inkplanner, park2023designing}  that satisfice the collage factor~\cite{buschek2024collage}.}

\change{Our work with \system builds on \citet{zhang2023visar}, supporting visual representation for hierarchical planning. It uses text fragmentation in the form of cells (or layers in our case) from \citet{kim2023cells}. It separates LLM content from human writing, per \citet{hoque2024hallmark}. Finally, it draws from the analytical, constructive, and critical aspects of collages \cite{buschek2024collage}. We also borrow the concept of LLM personas from ~\citet{benharrak2024writer} to improve human-centered writing. Together, these papers help us address the gaps in control and flexibility in LLM co-writing interfaces. \system uses a layered, zoomable workspace. It lets writers interact with LLMs in a non-linear, iterative way. This supports complex writing tasks and divergent thinking. It also keeps users in control of the content and structure.}
\section{\change{\system: A Layered Interface For LLM Co-Writing} }
\begin{figure*}[t!]
    \centering
    \includegraphics[width=\textwidth]{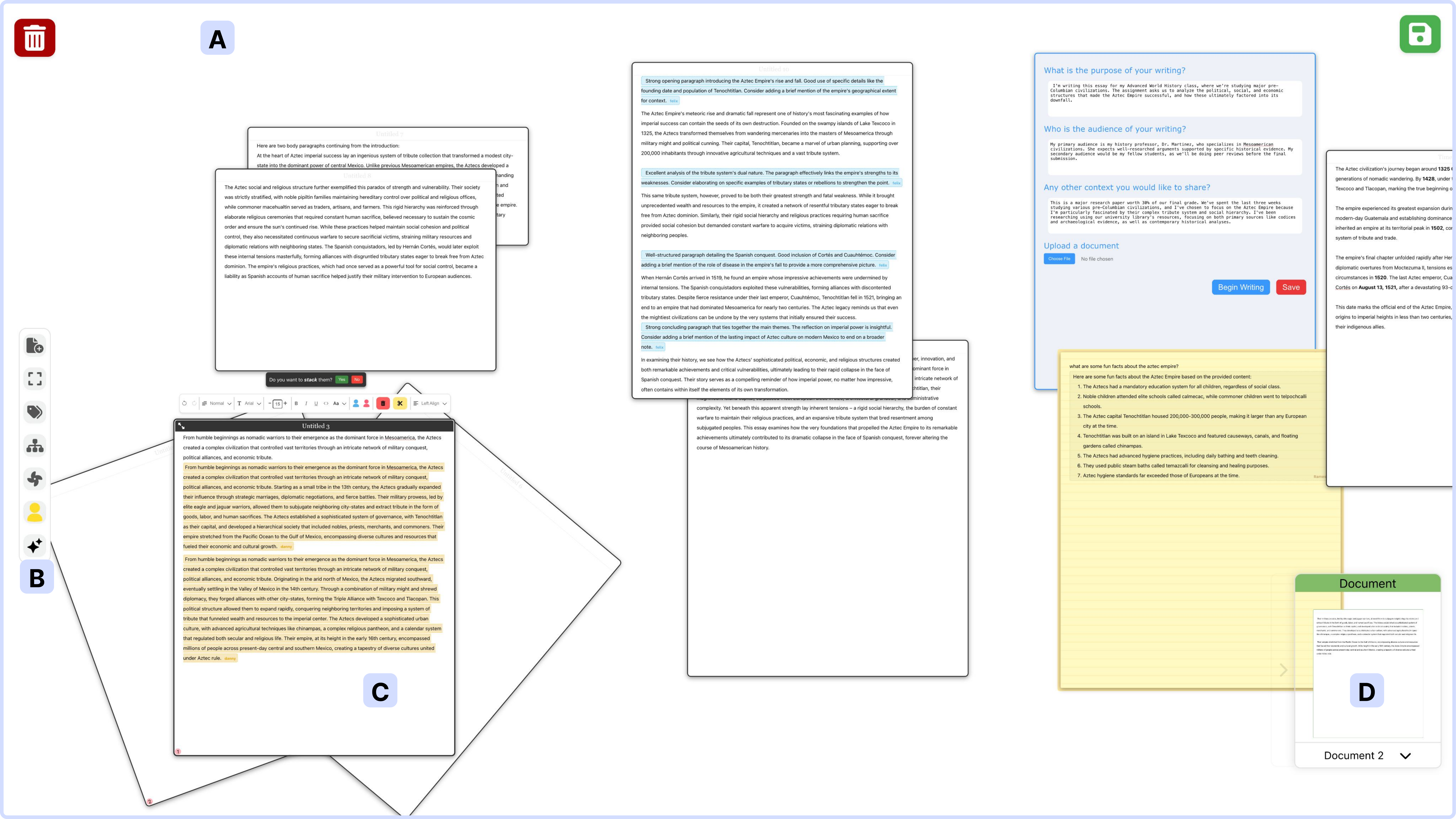}
    \caption{Workspace view of \system: (a) Zoomable,  scrollable writing workspace, (b) workspace level operations, (c) a text layer with layer toolbar on top, and (d) the compiled document viewer.}
    \label{fig:canvas}
\end{figure*}

\system is designed to support flexible generative writing workflows by interleaving content development and structural organization. In designing \system, our goal was to lower the \textit{envisioning gaps} around AI capabilities, instructions, and intentionality as well as to reduce the \textit{semantic} and \textit{articulatory} distance when interfacing with Generative AI. As shown in figure~\ref{fig:canvas}, the primary interface consists of an infinite zoomable \textit{workspace} where the writer can create and organize individual \textit{layers}. A \textit{layer} is a discrete, modular content unit within the writing workspace designed to encapsulate specific elements of the writing or organizational process. 

\change{Here, we use the ``layered'' metaphor because it aligns with how individuals think, by separating, refining, and integrating complex ideas. Unlike layers in visual tools such as Photoshop or Illustrator, which focus on the compositional blending of visual elements, \systems layers emphasize cognitive composition, representing distinct aspects of the writing or organizational process—such as brainstorming, tone adjustments, and meta-information. This metaphor naturally accommodates iterative workflows, enabling users to ``tear apart,'' ``combine,'' or ``stack'' layers for experimentation, similar to rearranging layers on a desk. Each layer acts as a \textit{sandbox} for specific tasks, preserving creative control and enabling non-linear exploration. By reimagining the familiar concept of layers, \system bridges user expectations with the flexibility and modularity required for writing tasks, where the process of exploration and refinement is as critical as the outcome.}

\system supports three types of layers, which include (1) a templated metadata layer for specifying the overall writing goals, (2) content layers for authoring the main text, and (3) a scratchpad for gathering relevant information to support the writing process. \system also supports `tags' for labeling individual layers and collections of layers (e.g., a stack of layers). Further, the interface provides contextual toolbars at the layer and workspace level. The layer-level toolbar consists of standard text formatting operations and specific AI functionality. The workspace-level toolbar supports adding layers and intelligent structuring operations. Further, \system implements an extensible collection of AI ``friends'' that provide targeted writing support for ideation, content organization, research, adjusting the tone, etc. To better understand how \system supports writing, let us follow Sashi, a freelance writer whose goal is to write an opinion piece about ``LLMs and Creative Ownership.'' 

\begin{figure*}[htbp!]
    \centering
    \begin{subfigure}[b]{0.28\textwidth}
        \centering
        \includegraphics[width=\textwidth]{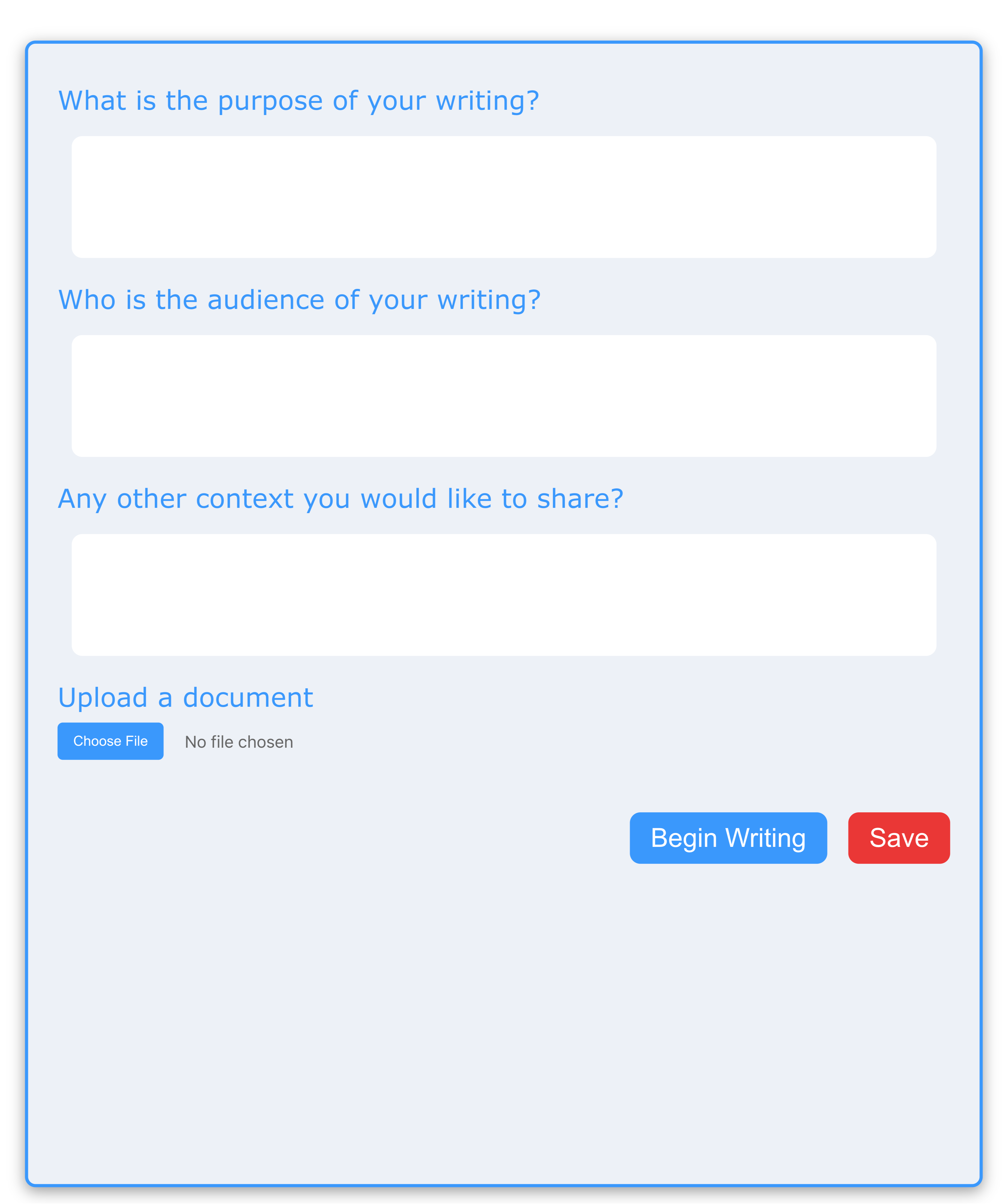}
        \caption{Specifying context of writing in meta layer}
    \end{subfigure}
    \hfill
    \begin{subfigure}[b]{0.28\textwidth}
        \centering
        \includegraphics[width=\textwidth]{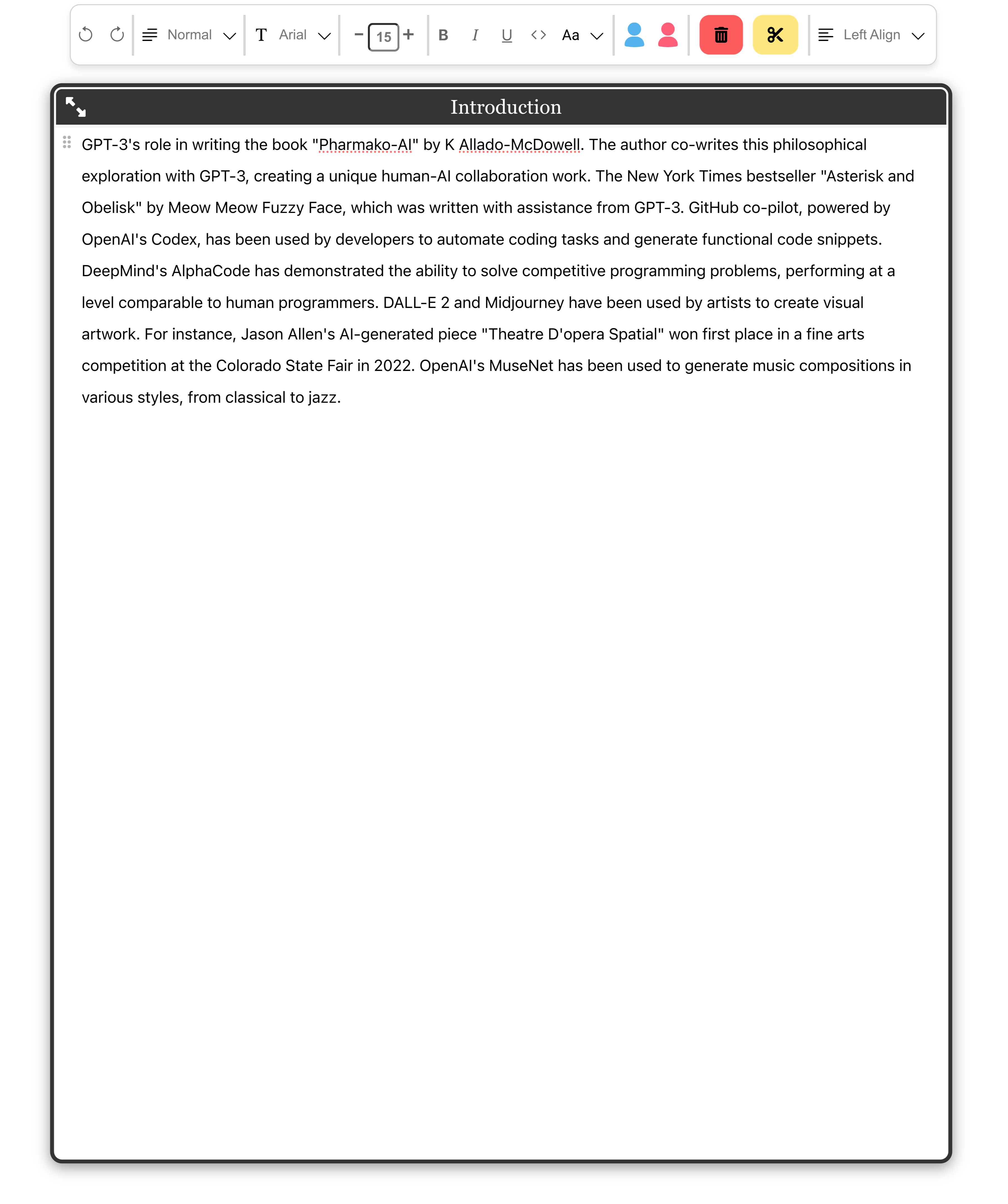}
        \caption{Starting free writing in Introduction Layer}
    \end{subfigure}
    \hfill
    \begin{subfigure}[b]{0.28\textwidth}
        \centering
        \includegraphics[width=\textwidth]{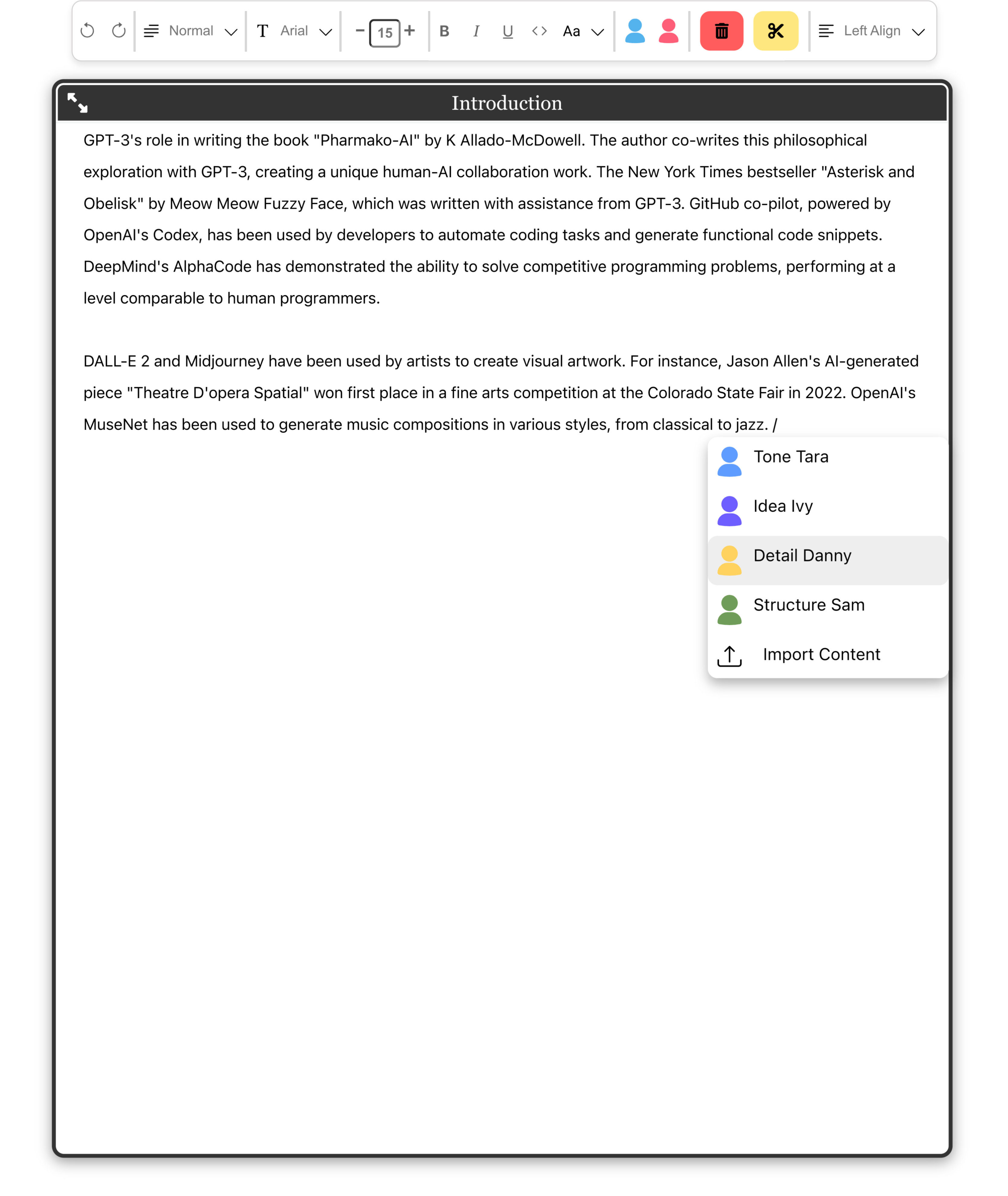}
        \caption{Calling Detail Danny from the `/' trigger menu}
    \end{subfigure}
    
    \vspace{1em}
    
    \begin{subfigure}[b]{0.28\textwidth}
        \centering
        \includegraphics[width=\textwidth]{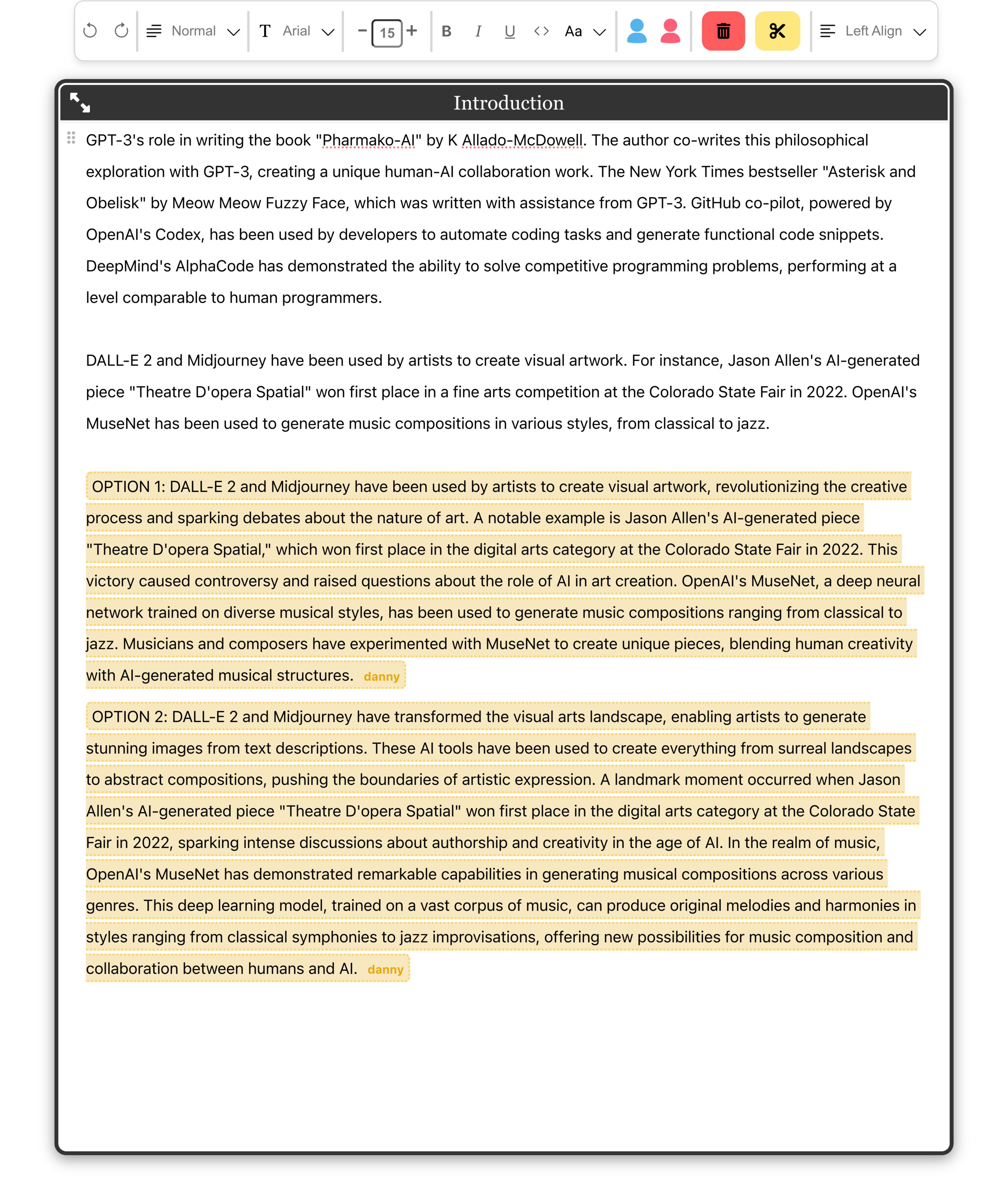}
        \caption{Detail Danny generating two variations for user-specified prompt for detail/elaboration}
    \end{subfigure}
    \hfill
    \begin{subfigure}[b]{0.28\textwidth}
        \centering
        \includegraphics[width=\textwidth]{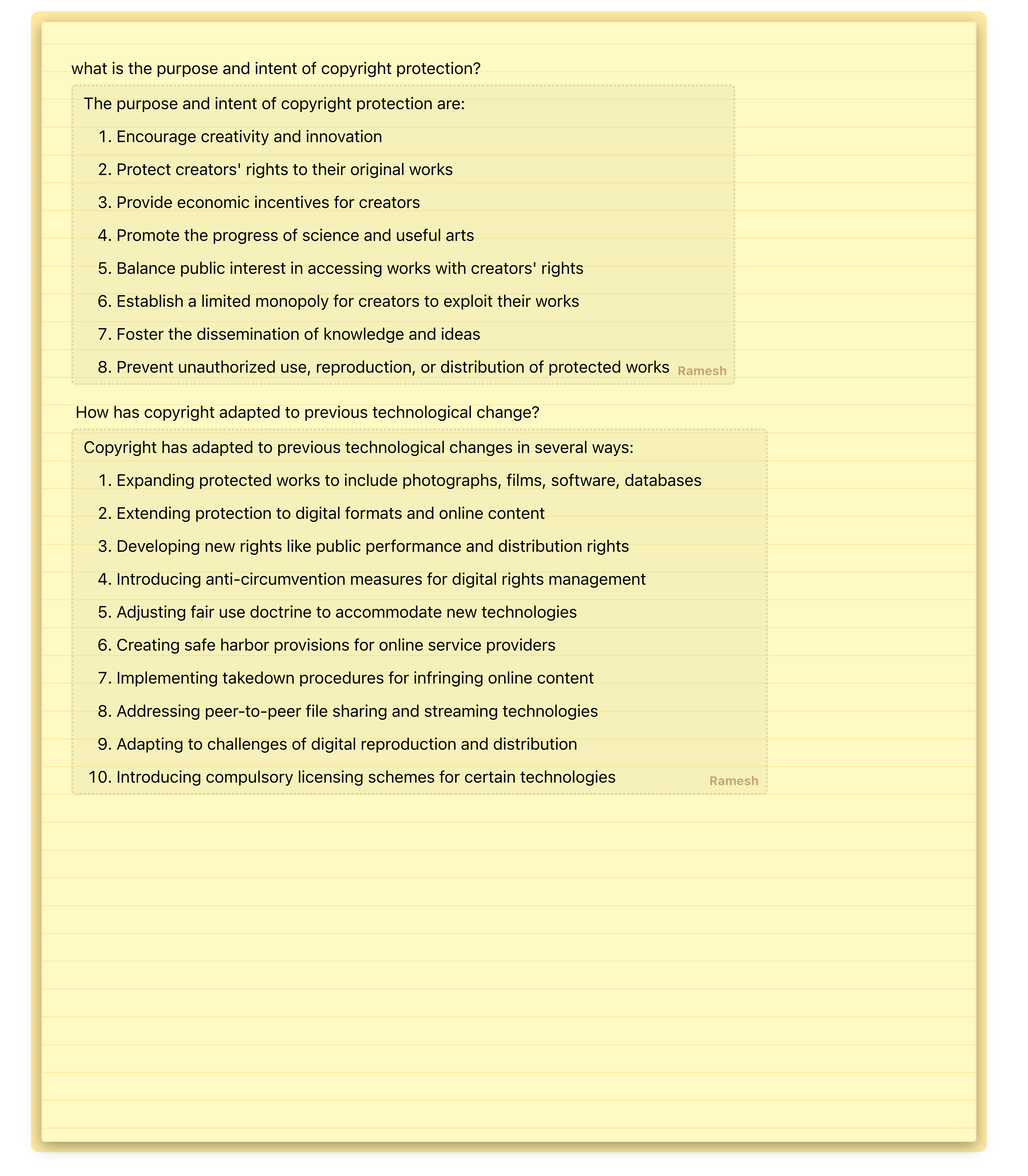}
        \caption{Research friend as a scratchpad for researching topics specified in meta layer}
    \end{subfigure}
    \hfill
    \begin{subfigure}[b]{0.28\textwidth}
        \centering
        \includegraphics[width=\textwidth]{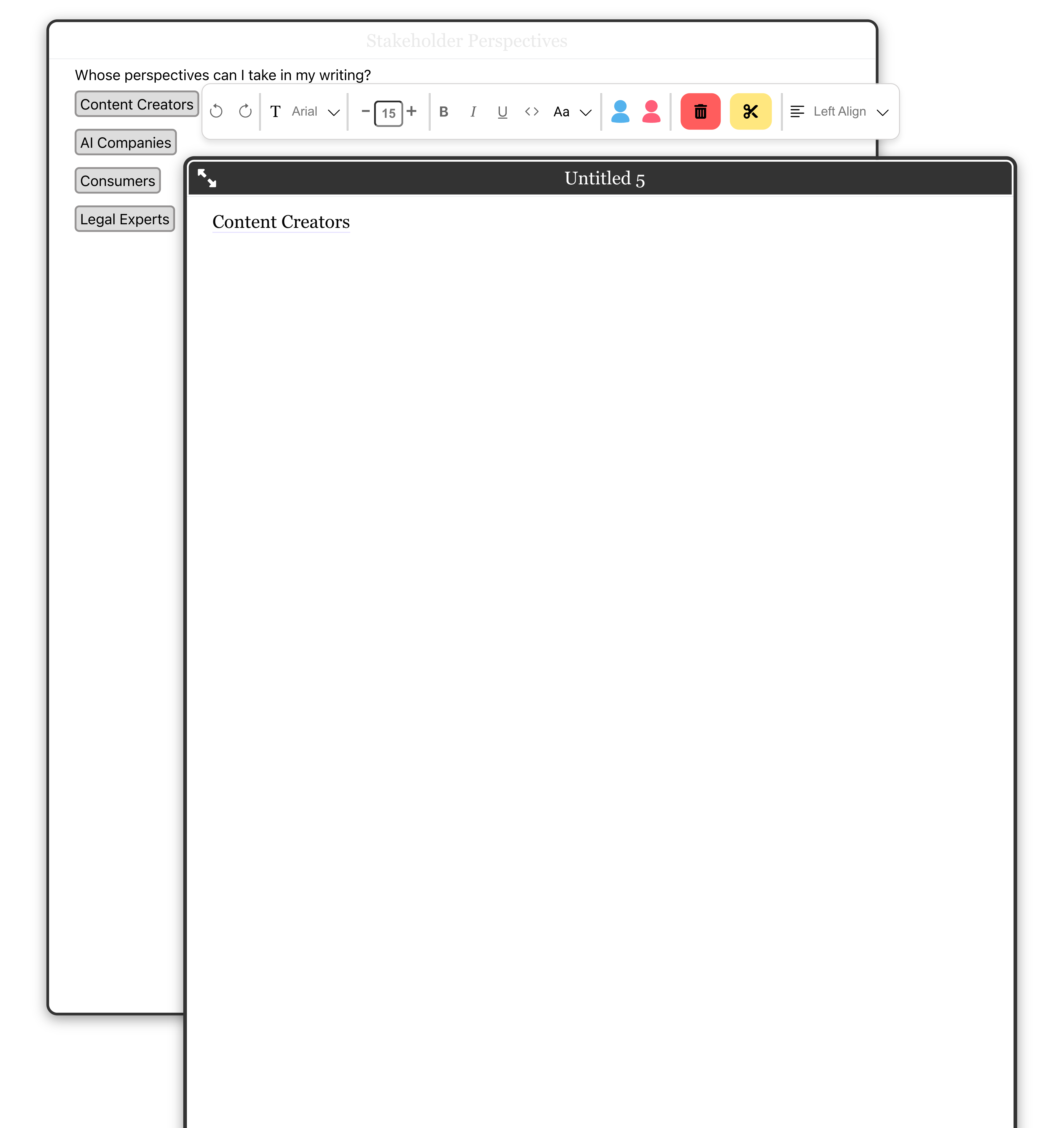}
        \caption{Adding links for child layers that the current layer will extract content from}
    \end{subfigure}
    
    \vspace{1em}
    
    \begin{subfigure}[b]{0.28\textwidth}
        \centering
        \includegraphics[width=\textwidth]{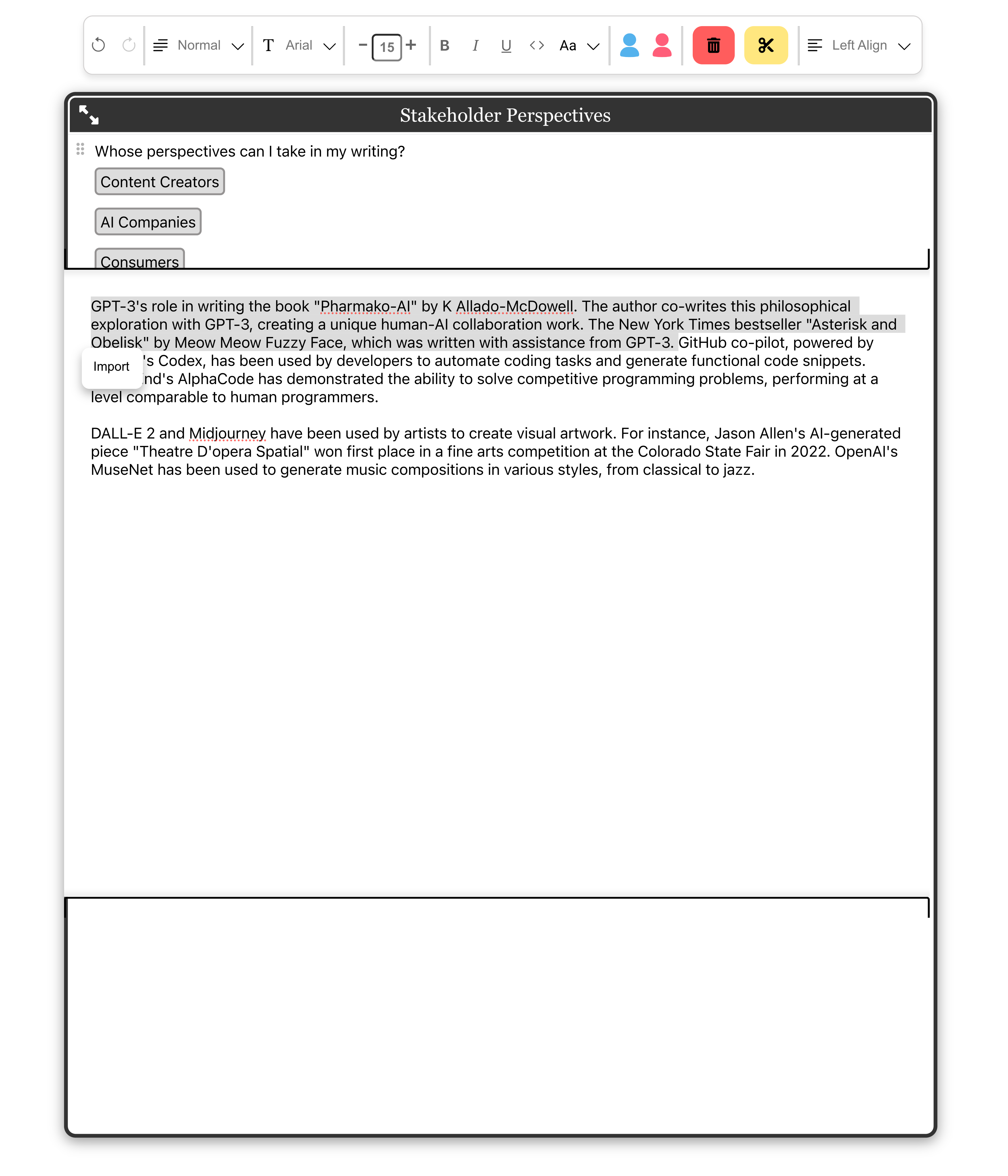}
        \caption{Tunneling into content from another layer into current layer}
    \end{subfigure}
    \hfill
    \begin{subfigure}[b]{0.28\textwidth}
        \centering
        \includegraphics[width=\textwidth]{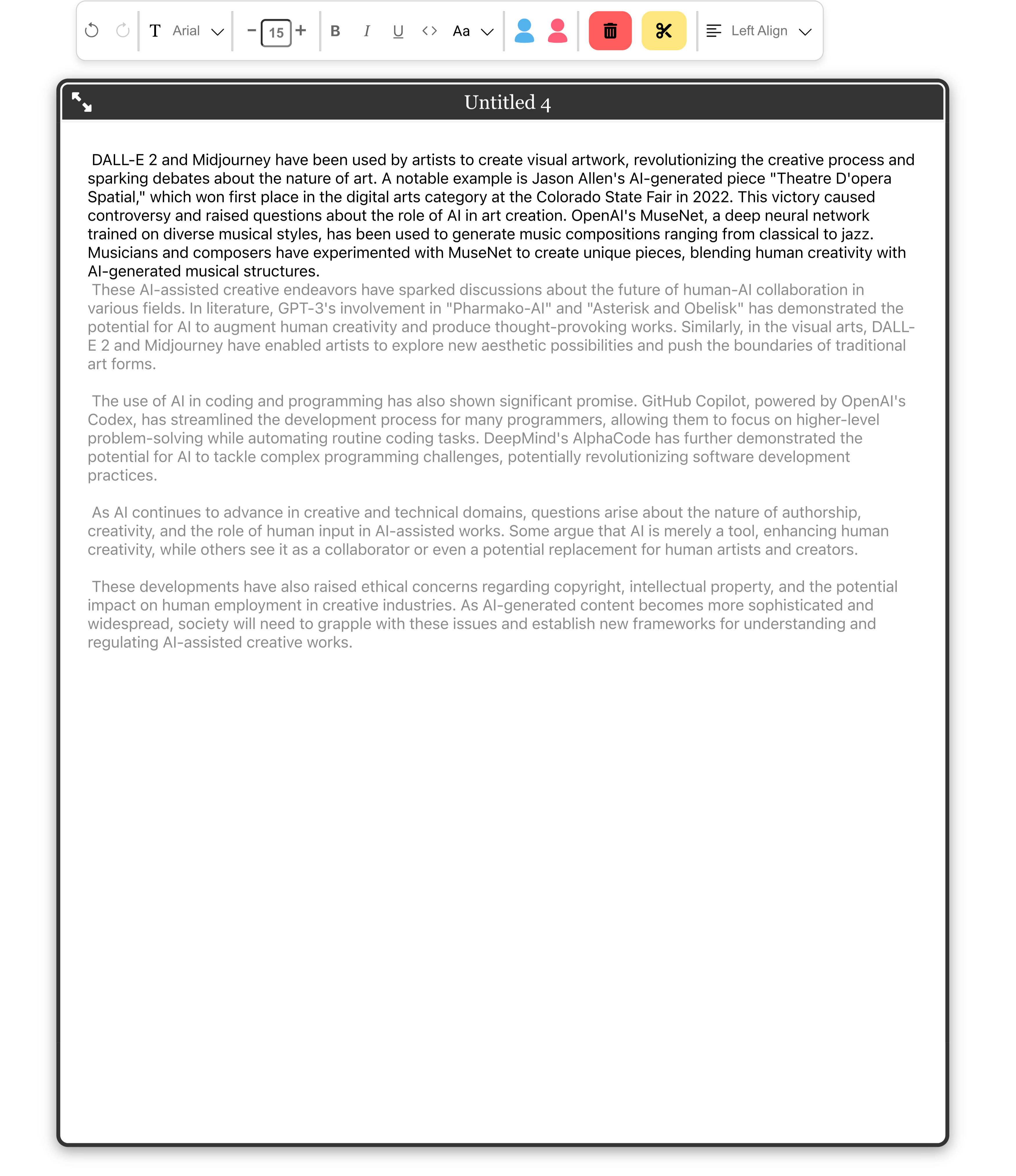}
        \caption{Peaking at (possible) future content using the bottom right}
    \end{subfigure}
    \hfill
    \begin{subfigure}[b]{0.28\textwidth}
        \centering
        \includegraphics[width=\textwidth]{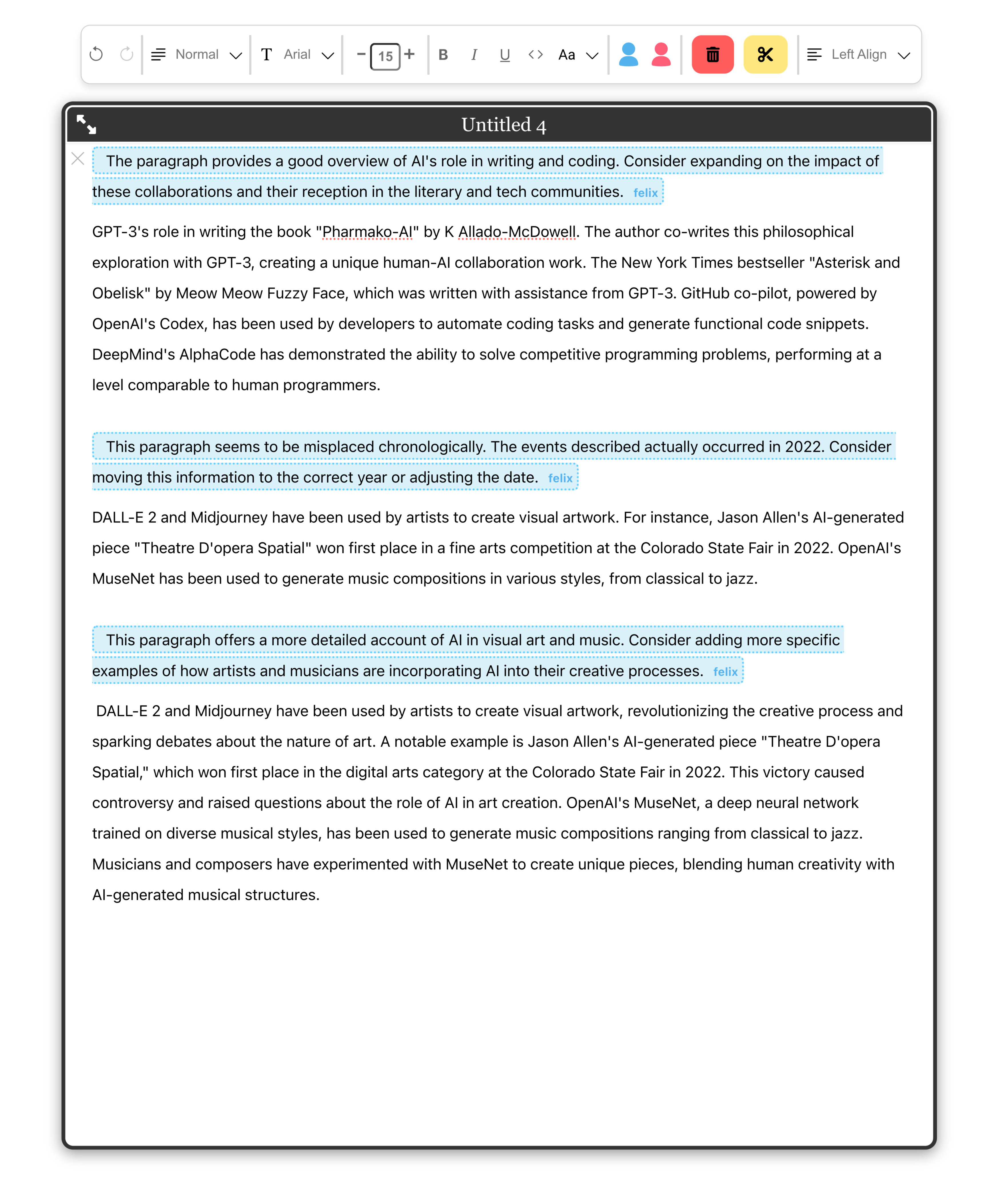}
        \caption{Invoking Feedback Felix for paragraph level feedback}
    \end{subfigure}
    \caption{Key Interactions supported by \system.}
    \label{fig:nine_figures}
\end{figure*}

\change{\subsection{Use Scenario}}
\subsubsection{Goal Setting}
To begin with, Sashi opens \system on his web browser and creates a new project. \system displays the workspace, adding a blank metadata layer~\textcolor{skyblue}{\faFile} to it (Figure~\ref{fig:nine_figures}a). The \textbf{metadata layer} contains a set of guiding questions and response text fields for writing goal, audience, topic context, and intent for writing. The metadata layer optionally allows Sashi to upload relevant documents to the writing task. In this case, Sashi enters the goal as \textit{``to write an opinion piece that explores the complex interplay between Large Language Models (LLMs), creative processes, and copyright law,''} the target audience as ``technology creators and potentially legal professionals,''  and intent as ``arguing for a reevaluation of current copyright frameworks to address the challenges posed by AI-generated content.'' Here, Sashi can also upload relevant documents, such as the news article about the AI-generated photo that won the art prize at the Colorado State Fair\footnote{\url{https://www.nytimes.com/2022/09/02/technology/ai-artificial-intelligence-artists.html}}, the New York Times Lawsuit against OpenAI\footnote{\url{https://nytco-assets.nytimes.com/2023/12/NYT_Complaint_Dec2023.pdf}}, and any other relevant legal documents. Sashi can always revisit this layer throughout his writing to provide additional context. At this point, Sashi clicks on the ``Begin Writing'' button at the bottom of the metadata layer. This adds a new layer to the workspace, which Sashi names `Introduction'  \pageiconwithnumber[lightestgray]{black}{I}.

\subsubsection{Content Development}
On this layer \pageiconwithnumber[lightestgray]{black}{I}, Sashi begins by engaging in \textit{free-writing}, exploring various real-world instances of LLM creativity he has encountered to motivate the article (Figure~\ref{fig:nine_figures}b). Once he has written several examples, he realizes that he must also include a brief description of LLMs and their creative abilities. Instead of manually describing LLMs, Sashi opts to leverage the content development features of \system. Inspired by `personas' for writing feedback~\cite{benharrak2024writer}, we have implemented a set of \textbf{writer's friends} that embody various writing skills, including \coloredcircle{IdeaIvy} Idea Ivy for content ideation, \coloredcircle{DetailDanny} Detail Danny for elaboration, \coloredcircle{StructureSam} Structure Sam for content structuring, \coloredcircle{ToneTara} Tone Tara for stylistic suggestions, \coloredcircle{FeedbackFelix} Feedback Felix for feedback on content, and \coloredcircle{AudienceAli} Audience Ali for audience-specific feedback.

In this case, Sashi selects the Detail Danny friend by typing a forward slash `/' at the beginning of a new paragraph. This action opens a dropdown menu displaying a list of all friends, from which Sashi chooses \coloredcircle{DetailDanny} Detail Danny (Figure~\ref{fig:nine_figures}c). \system inserts an \customdashedbordertext{inline prompt box}{DetailDanny}{DetailDanny} where he writes \textit{``Write a concise description of LLMs and their creative abilities, explain how they function for the creative tasks listed above''} and presses the enter key. In response, \system generates a paragraph of text about LLMs and creativity. The \customsolidbordertext{generated text}{DetailDanny}{DetailDanny} is highlighted with a background color corresponding to the friend (Figure~\ref{fig:nine_figures}d). Sashi can accept this content by pressing enter at the end of the content or delete it and regenerate new content. Sashi accepts the suggestion and continues editing. Sashi has a high-level vision about a few topics his opinion essay should cover. He creates new layers for `Traditional Copyright Law - \pageiconwithnumber[lightestgray]{black}{L},' `Authorship -- \pageiconwithnumber[lightestgray]{black}{A},' and `Ethical Considerations -- \pageiconwithnumber[lightestgray]{black}{E}' and, as before, writes his thoughts. When writing about copyright law, Sashi realizes he needs to understand copyright principles better. He creates a new \textbf{scratchpad} layer ~\textcolor{legalyellow}{\faFile}, which is supported by Research Ramesh, where he can ask questions about the purpose and intent of copyright protection or how copyright has adapted to previous technological changes to help with his main content writing (Figure~\ref{fig:nine_figures}e). 

In the course of writing, Sashi creates a new layer for Stakeholder Perspectives -- \pageiconwithnumber[lightestgray]{black}{S} and brainstorms using  \coloredcircle{IdeaIvy} Idea Ivy about whose perspectives to include, and in response, Ivy suggests talking about \customsolidbordertext{Content Creators}{IdeaIvy}{IdeaIvy}, \customsolidbordertext{AI Companies}{IdeaIvy}{IdeaIvy}, \customsolidbordertext{Consumers}{IdeaIvy}{IdeaIvy}, and \customsolidbordertext{Legal Experts.}{IdeaIvy}{IdeaIvy}. By individually selecting each stakeholder, Sashi can create \textbf{sub-layers} to flesh out the stakeholder perspective (Figure~\ref{fig:nine_figures}f). There is a persistent link between the text in the Stakeholder layer \pageiconwithnumber[lightestgray]{black}{S} and the sub-layers [\pageiconwithnumber[lightestgray]{black}{$S^{CC}$}, \pageiconwithnumber[lightestgray]{black}{$S^{AI}$}, \pageiconwithnumber[lightestgray]{black}{$S^C$}, \pageiconwithnumber[lightestgray]{black}{$S^{LE}$}]. Sashi can also \textbf{split} the content into two layers and develop the content separately.  

Idea Ivy can also be used to support more complex writing procedures across layers, such as how some specific text about the economic impact in the Ethical Considerations layer \pageiconwithnumber[lightestgray]{black}{E} might relate to the current text in the content-creator layer \pageiconwithnumber[lightestgray]{black}{$S^{CC}$}. To support such complex prompting using content across layers, \system supports a \textit{tunneling} function where the current layer split opens at the cursor position to reveal a desired layer selected from the dropdown (Figure~\ref{fig:nine_figures}g). From here, Sashi can choose the relevant text and import it to the current prompt he is composing. On any layer, if he is stuck on what to write, he can click on the bottom right corner of the layer to \textbf{`peek'} into what the Generative AI might write (Figure~\ref{fig:nine_figures}h). The peek function shows continuing text based on his current writing in a greyed-out format. For creating structure from unstructured text, he can invoke \coloredcircle{StructureSam} Structure Sam, who will create a new layer with a structured representation of the content, including headings and subheadings. He can invoke the \coloredcircle{AudienceAli} audience and \coloredcircle{FeedbackFelix}feedback friends on any layer using the toolbar option to receive comments inline, which he can toggle on or off (Figure~\ref{fig:nine_figures}i). He can call upon \coloredcircle{ToneTara} Tone Tara to generate a different stylistic variations of the current content rendered as separate layers from the original. 

Lastly, as he is writing, Sashi can \textbf{compare} two layers by bringing them close, such that the right edge of a layer is touching the left edge of the other layer \pageiconwithnumber[lightestgray]{black}{$S^{AI}$}\pageiconwithnumber[lightestgray]{black}{$S^{CC}$}. \system recognizes this as an intent to compare and provides a floating button, ``Compare the two layers?'' On clicking this, the button expands into a text prompt where Sashi can input specific instructions for comparison, such as ``how the stakeholder perspectives of content creators and AI companies align or conflict regarding LLM-generated content and copyright.'' The system uses color highlighting and inline annotations to indicate similarities and differences. To mitigate possible confusions, these annotations only persist while the layers remain in proximity. 

\subsubsection{Structure and Rhetoric}
\begin{figure*}[t!]
    \centering
    \includegraphics[width=0.8\textwidth]{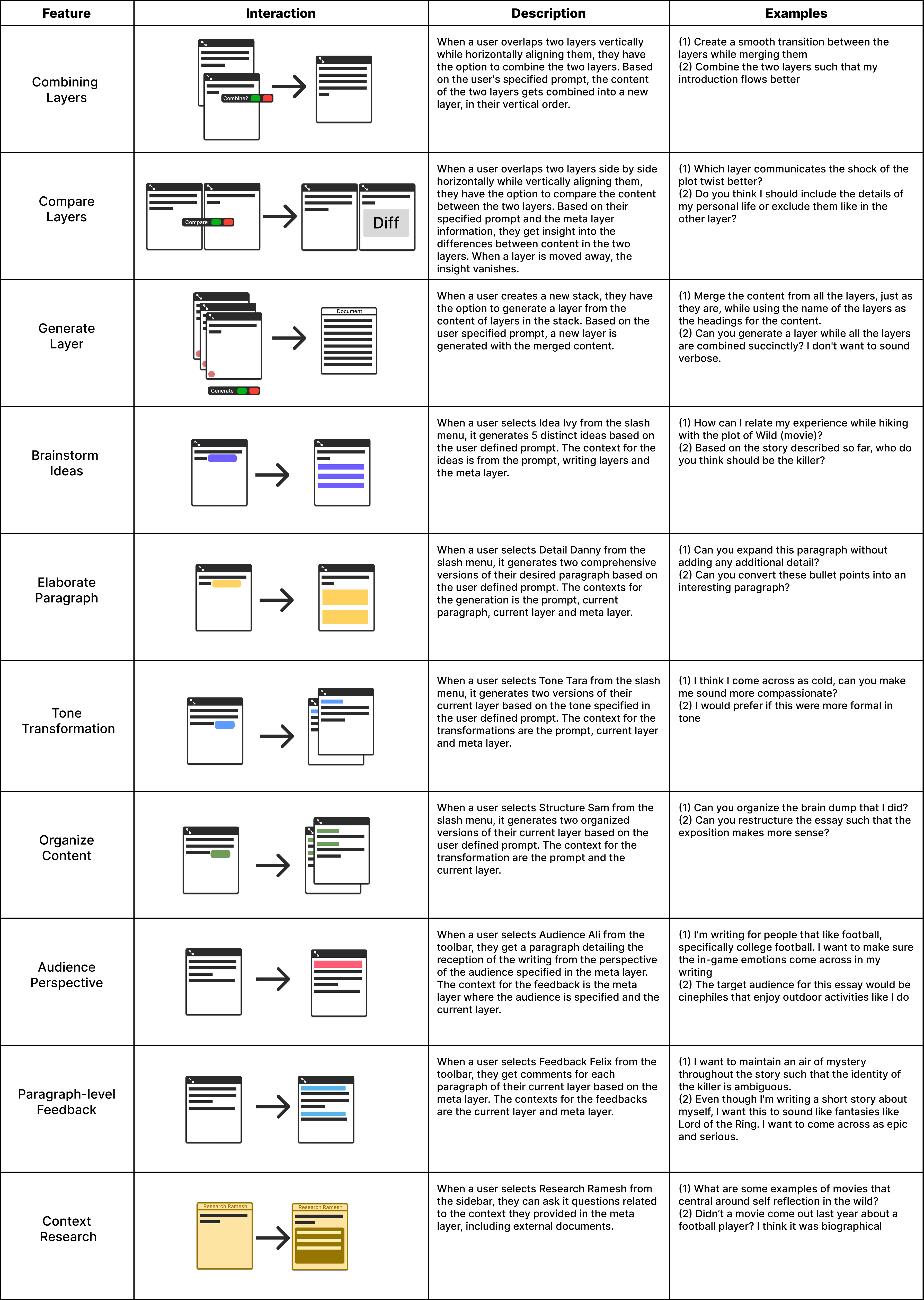}
    \caption{\change{Content transformation through user prompts. The following 10 features allow writers to issue descriptive instructions for invoking specialized LLM assistance. The example column consists of real prompts issued by participants in the usability assessment.}}
    \label{fig:layered_action}
\end{figure*}

Throughout this process, \system provided several features for Sashi to develop the structure for his final essay and tailor it to specific audiences. To organize the content, \system supports \textbf{combining} two layers into a single layer by bringing the second layer to the bottom of the first layer \twopageicon{lightestgray}{lightestgray} and optionally prompting for specific transitional text between layers. For instance, he can connect a layer on `Challenges posed by LLMs' \pageiconwithnumber[lightestgray]{black}{Ch} with the layer on `Copyright Law' \pageiconwithnumber[lightestgray]{black}{C} by generating transitional text from identifying the problems to analyzing the legal framework meant to address them. Alternately, Sashi can \textbf{tear} layers into parts and glue them together to reorganize the text, i.e. to have content inform structure. Sashi can fold layers he wishes to exclude, which show the text summary on the folded side of the layer. He can \colorbox{yellow}{Tag} individual layers or even clusters and stacks of layers in the workspace. When Sashi is ready to \textit{compose} his essay, he can \textbf{stack} the layers manually in an order [\pageiconwithnumber[lightestgray]{black}{I},\pageiconwithnumber[lightestgray]{black}{C},\pageiconwithnumber[lightestgray]{black}{E},\pageiconwithnumber[lightestgray]{black}{S}\ldots] and generate the final document by concatenating them or asking the LLM to generate the order based on an unordered stack. Here he can issue specific prompts such as generating an audience-specific version, editing for consistency across layers, or even adaptive summarization to meet some target length. The final document is generated with any edited text highlighted for Sashi to review. Sashi can click on the text in the final document and track back to the source layer. In this manner, \systems features allow Sashi to use generative AI for content development and rhetorical strategies, flexibly, fluidly, and iteratively. Table~\ref{fig:layered_action} summarizes the different scripting and shifting affordances of \system.


\section{System Architecture}


\begin{figure*}[t]
    \centering
    \includegraphics[width=\textwidth, height=0.55\textheight, keepaspectratio]{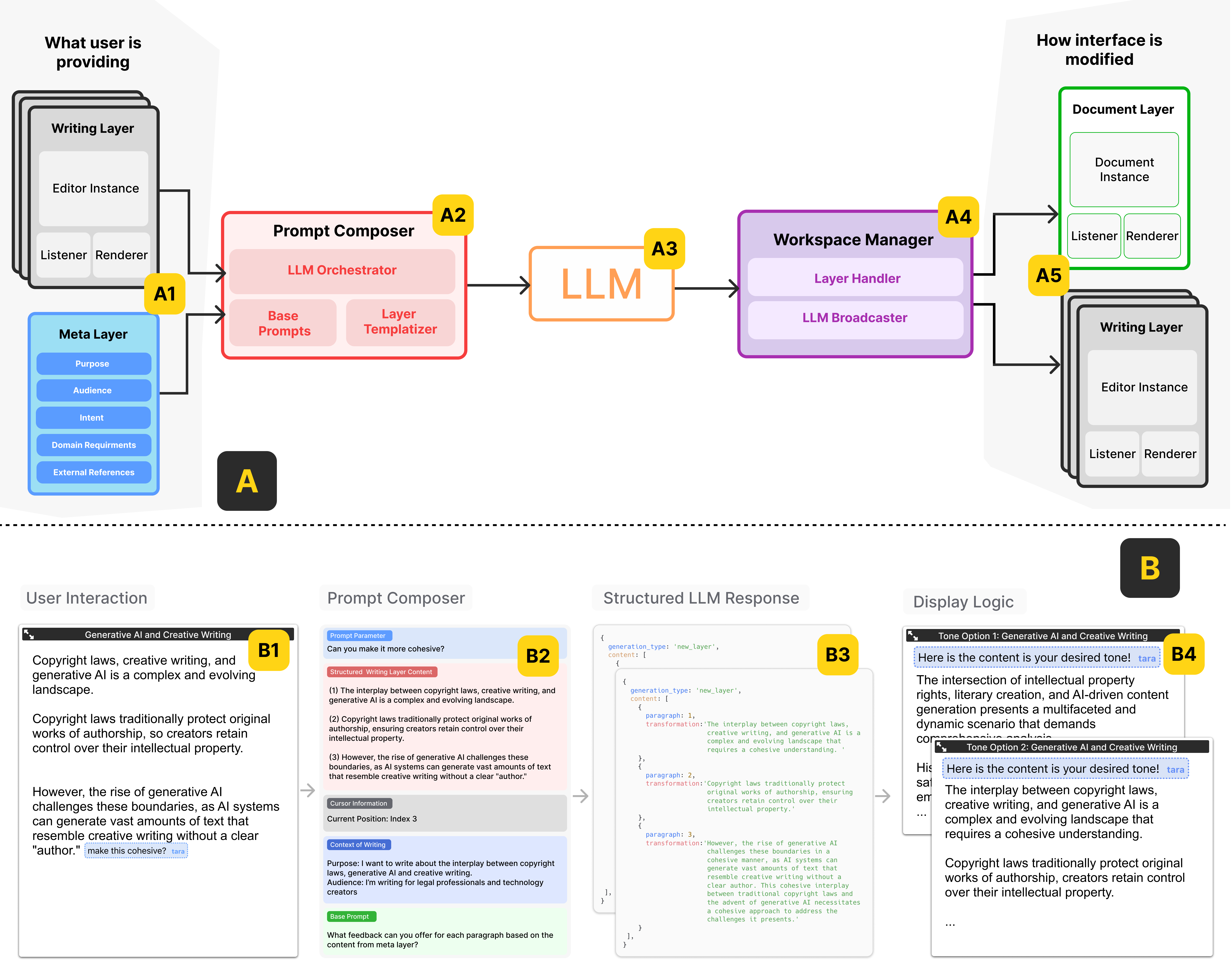}
    \caption{\change{(A) \textbf{system architecture} and (B) \textbf{example} of how it is leveraged for Tone Tara. \textbf{(A1\&B1)} The meta layer and writing layers provide the content, prompts and other details for prompt composer. \textbf{(A2\&B2)} The prompt composer uses LLM Organizer and base prompts to structure the context and send the details for generation constraints using Layer Templatizer. \textbf{(A3\&B3)} LLM outputs the structured responses. \textbf{(A4\&B4)} The Workspace manager projects the response onto writing layers and document layers using the layer handler and LLM broadcaster.}}
    \label{fig:system_architecture}
\end{figure*}

\change{At a high level, \system is a tool for generating text from input across multiple text editors and displaying it in multitudinous ways, depending on writer purposes.  The generated text can be rendered as a highlighted block at targeted positions in existing text editors, or it can be rendered in a newly instantiated layer. To support these operations, as shown in Fig. ~\ref{fig:system_architecture} A, the system consists of (1) layer primitives that encapsulate various kinds of editors (Fig. ~\ref{fig:system_architecture} A1); (2) the prompt composer for converting the input for generation into well-defined structures (Fig. ~\ref{fig:system_architecture} A2); (3) the large language model for text generation (Fig. ~\ref{fig:system_architecture} A3); and (4) the workspace manager that controls the layer primitives and associated display logic (Fig. ~\ref{fig:system_architecture} A4). Here, we describe the core functionality of each module through an example (Fig. ~\ref{fig:system_architecture}} B) of \textit{Tone Tara}.

\subsection{Layer Primitives}
To support the layered interface paradigm, \system implements a number of layer primitives.

\subsubsection{Writing Layer}

\change{The Writing Layer is the primary interface component that supports all content-related operations. Internally, this layer consists of three key modules: an editor instance, a listener module, and a text renderer. The editor instance, which is what the writer interacts with, provides standard text editing capabilities, and the listener module monitors and subscribes to LLM outputs, ensuring that any AI-generated content or suggestions are dynamically integrated into the writing process. Finally, the text renderer handles the interaction and display, such as invoking specific prompt composition interactions and displaying the LLM's outputs in the editor. Internally, the text editor is maintained as a dynamic text-based template with user-written text, AI-provided input, and placeholder fields to display AI content. Placeholder fields are dynamically added during LLM interactions and serve as areas where AI-generated content can be inserted, modified, or replaced. The blue text boxes in Fig. ~\ref{fig:system_architecture} B4 are examples of placeholder fields.}

\subsubsection{Meta Layer} The Meta Layer \change{(blue box in Fig. ~\ref{fig:system_architecture} A1)} provides essential context for the current workspace, acting as a foundational layer that informs all LLM inputs and outputs (i.e., the global context). Information added to this layer, such as writing goals, target audience, key themes, and specific instructions, serves as a user-defined base instruction for the system. By embedding these instructions into the workspace, the Meta Layer allows the system to consistently generate outputs that adhere to the user’s intent, making the writing process more cohesive and context-sensitive. Internally, this is a JSON object with attributes for (1) purpose, (2) audience, (3) intent, (4) domain requirements, and (5) external references, which is a list of supporting documents or resources relevant to the writing task. 

\subsubsection{Document Layer} The Document Layer represents the final `compiled' content from a set of layers that can be specified by the writer. The Document Layer’s non-editable format preserves LLM contributions, ensuring structural integrity and preventing accidental changes. It provides a clear separation between drafting and the finalized document. Unlike the Writing Layer’s editor instance, the Document Layer uses a document instance—a JSON object that stores an index and editor content—allowing the final output to be rendered and preserved as a cohesive whole. The Document Layer incorporates hyper-refs that connect specific sections of the final output back to their corresponding Writing Layers, enabling writers to trace content origins, review edits, and maintain a clear understanding of how each section evolved.

\subsection{Prompt Composer}
The prompt composer \change{(Fig. ~\ref{fig:system_architecture} A2)} plays a central role in relaying interactions between the users and the LLM. It is made up of three key components: System Prompts for various writing tasks, Layer Templatizer, and LLM Orchestrator. The System Prompts and template logic are stored in a task knowledge dictionary, which functions as a repository of predefined tasks, each with its own associated prompts and processing rules. The LLM Orchestrator integrates inputs from Writing Layers, the Meta Layer, prompt parameters (user-defined prompt instructions in real-time interactions), and task knowledge. 

When the user carries out an interaction, such as requesting assistance with elaboration or restructuring, the Prompt Composer first retrieves the task-specific knowledge from the repository and sends it to the Layer Templatizer. The Templatizer then parses the Writing Layer’s context to determine the relevant position in the document where the LLM content should be rendered. For example, if the user selects a paragraph for rewording, the Templatizer identifies the paragraph's location in the writing layer, updates the template with the task knowledge, and specifies where the LLM's output (e.g., the reworded text) will appear. Lastly, the LLM Orchestrator handles the concatenation order of prompt parameters, structured content, and base prompts. It also determines the number of LLM instances needed and manages how the information is processed and passed to the LLM, ensuring smooth integration of all inputs. In case of \textit{Tone Tara}, the composed prompt after LLM Orchestrator is applied, will look like Fig. ~\ref{fig:system_architecture} B2.

\subsection{Large Language Model}
The LLM component in our system leverages the Claude 3.5 Sonnet~\cite{anthropicclaude} model from Anthropic, selected for its strong language comprehension, ability to handle complex writing tasks, and efficiency in generating high-quality creative outputs. Following the \change{Layer Templatizer}'s instructions from the Prompt Composer, the model is instantiated to generate either templatized or free-form text, depending on the requirements of the task. The LLM Component operates in close coordination with the Prompt Composer, processing the consolidated input and producing outputs that align with the context and specific objectives of the user’s writing task. \change{We can see this in action in Fig. ~\ref{fig:system_architecture} B3 where the LLM generates two structured responses, each for a new layer to be created}.

\subsection{Workspace Manager}
\change{The Workspace Manager controls the flow of LLM-generated content through two tightly coupled subsystems.}

\subsubsection{\change{Layer Handler}} \change{The Layer Handler maintains the structural integrity of the writing workspace through nested array-based data structures, managing layer operations like stacking, folding, fanning, and comparing. It handles the creation of new layers when requested by the Layer Templatizer and maintains spatial arrangements and parent-child relationships.}

\subsubsection{LLM \change{Broadcaster}} The \change{LLM Broadcaster} is responsible for managing the flow and distribution of LLM-generated content across layers, ensuring that AI outputs are properly integrated into the writing process. When the LLM generates content, the \change{Writing} Layer's listener module subscribes to these outputs, dynamically capturing and processing them. 

\change{Together, these subsystems ensure seamless integration of AI-generated content while maintaining the structural coherence of the writing workspace.}

\subsection{Implementation}
To implement this framework in our interface, we built the application using React~\cite{react}. For the rich-text editor, we integrated Lexical~\cite{lexical}, which provides a hierarchical, node-based WYSIWYG editor with customizable listeners, allowing us to capture and process user inputs dynamically. To support the infinite canvas for spatial organization of content, we used React Flow~\cite{reactflow}, a library that enables the creation of interactive, movable components, allowing users to visually arrange and connect different writing elements. On the backend, we utilize Anthropic's Claude 3.5 Sonnet model~\cite{anthropicclaude} for LLM functionality. The application is deployed on Heroku~\cite{heroku}, and Firebase’s NoSQL cloud real-time database~\cite{firebase} manages the storage of user data. Additionally, we log important metrics, including user-defined prompts, feature invocation timestamps, and words per minute, to track interaction data and performance.

\section{Example Writing Workflows with \system}

In this section, we demonstrate the flexibility of writing with \system through three distinct workflows: freewriting, document-based-question, and parallel topic development. Each example highlights how the layered interface and AI tools support fluid, iterative writing across different starting points and content structures.

\subsection{Freewriting to Argumentative Writing Workflow}
In this workflow, the writer begins with freewriting, allowing ideas to flow without worrying about organization or structure. Once they have written their thoughts, they determine that an argument style structure can be effective to organize the text, consisting of the following components: Claim, Grounds, Warrant, Backing, Qualifier, and Rebuttal~\cite{toulmin2003uses,marshall1989representing}. The writer adds the \textit{argument template}  to \systems workspace and drops the layer with the freewriting text onto the template. Using the template, \system takes the unstructured text and generates six layers, one for each component. Next, the writer can continue to flesh out each layer using Writer's Friends such as ``Idea Ivy'' and later refine, reorganize, and link ideas across layers, transforming initial thoughts into a structured argumentative essay.

\begin{figure*}[t]
    \centering
    \includegraphics[width=\textwidth, height=0.3\textheight, keepaspectratio]{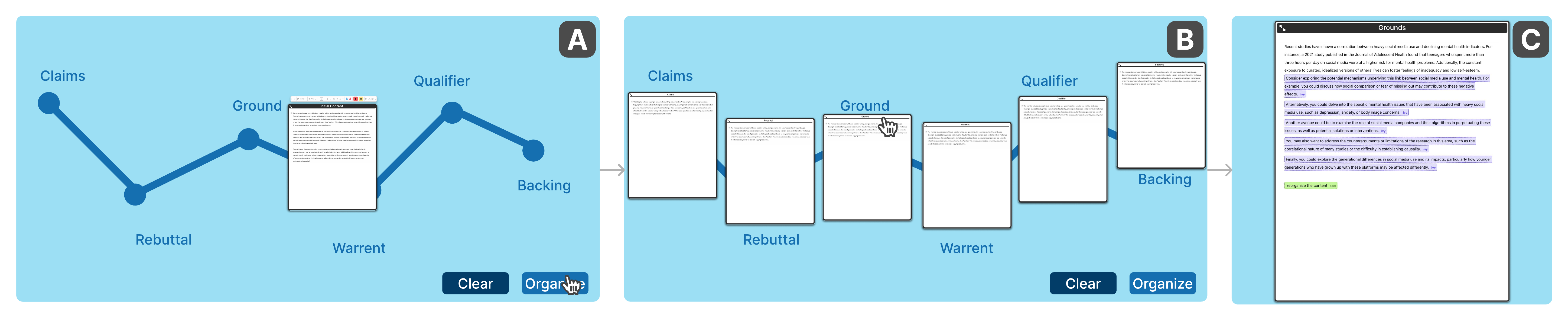}
    \caption{Free Writing Example with \system \change{ \textbf{(A)} Writer calls \textit{Template} for organizing their writing into milestones for argumentative writing. \textbf{(B)} They select the ``Ground'' layer for further development. \textbf{(C)} They invoke \textit{Idea Ivy} to brainstorm what to write next and then use \textit{Structure Sam} to organize that into subheadings and paragraphs.}}
    \label{fig:freewriting}
\end{figure*}

\begin{figure*}[t]
    \centering
    \includegraphics[width=\textwidth, height=0.3\textheight, keepaspectratio]{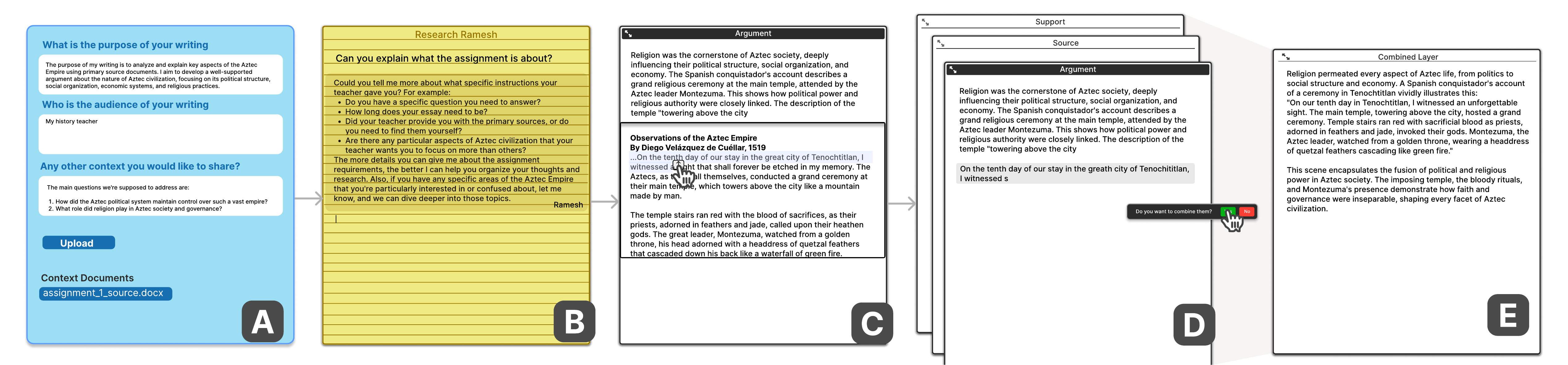}
    \caption{Document-Based Question Example with \system \change{ \textbf{(A)} The writer specifies the context of their writing and upload their assignment to the \textit{Meta Layer}. \textbf{(B)} They call \textit{Research Ramesh} to understand details of their assignment based on the context document. \textbf{(C)} They \textit{Tunnel} into another layer to extract some details. \textbf{(D)} They combine a stack of layers to generate \textbf{E}.}}
    \label{fig:freewriting}
\end{figure*}

\subsection{DBQ Writing Workflow for Students}
In this workflow, the student working on a Document-Based Question (DBQ) can upload primary source documents directly into the system. The metadata page allows the student to provide information about the assignment goals, guiding questions, and document context. The student can then create additional layers for organizing their arguments, categorizing evidence drawn from the primary sources. For example, they might have separate layers for economic, political, and social factors. LLM friends like ``Research Ramesh'' assist by analyzing the documents and suggesting relevant excerpts or summaries that will fit the argument. As the student develops their essay, cross-layer interactions allow them to pull evidence from the source layers into argument layers, ensuring that evidence is seamlessly linked back to the original documents. Using \systems stacking feature, the student can then combine the various layers into a cohesive essay. \system allows them to specify prompts for transitions in order to ensure that all arguments are well-supported by evidence. This workflow encourages students to engage deeply with primary sources, while providing an intuitive way to structure their argument, making the ballistic process of evidence-based writing more fluid and interactive.

\subsection{Layered Topic Development for Research Paper}

In this workflow, the writer develops different sections of the research paper in parallel, creating distinct layers for the literature review, methodology, and findings. \system allows the writer to work on these topics simultaneously, using cross-layer interactions to draw connections between sections. For instance, the writer can link methodology details to key studies in the literature review. Within the findings section, the writer can create separate layers for different findings or topics, allowing them to focus on each result individually. Once these layers are complete, the writer can combine them into a unified findings section, ensuring a cohesive and structured presentation of the results. Writer's friends like ``Research Ramesh'' assist by sourcing and summarizing relevant research, while ``Tone Tara'' and ``Feedback Felix'' help ensure stylistic and argumentative consistency across sections. Once the layers are fully developed, the document structuring features in \system help weave them together into a cohesive research paper.
\section{Evaluation}

To evaluate the effectiveness of our layered interface paradigm and generative authoring workflow, we conducted two studies: (1) a mixed-methods user experience evaluation with a Subjective Evidence-Based Ethnography (SEBE) protocol, and (2) a between-subjects deployment study on Prolific.

\subsection{User Experience Evaluation}

In the first study, our goal was to evaluate \system's impact on cognitive load, usability, and its influence on creative output. \change{ We wanted to understand the manners in which personas and layered affordances could bridge the envisioning gap~\cite{subramonyam2024bridging} for writers. }

\subsubsection{Participants}
We recruited participants via LinkedIn and Twitter, selecting individuals who engaged in creative writing at least a few times per month. The final sample included 12 participants (8 male, 4 female), aged 18 to 54. The majority (7 participants) were in the 18-24 age group, with 3 in the 25-34 range, and one each in the 35-44 and 45-54 groups. All participants were native English speakers. Educational backgrounds varied: 6 participants held bachelor's degrees, 5 had master's degrees, and 1 had some college experience. Participants reported diverse creative writing habits and experiences, with most having taken creative writing courses. Self-reported confidence in writing ranged from 2 to 5 on a 5-point scale ($\mu = 3.54$, $\sigma = 0.82$).

\subsubsection{Study Procedure}
Each study session lasted not more than 2 hours. At the start of the session, participants received a live demonstration of the application and were encouraged to ask questions. The participants were then asked to access the application on their own computer through the web url and complete a brief guided writing task about a tree, a cat and a dog to familiarize themselves with the interface. Next, participants engaged in the main writing task. They were given 40 minutes to write an 800-word essay connecting a recent real-life event to a film, using one of the 5 specified writing styles (First-Person Narrative, Journalistic Style, Dialogue Format, Letter or Diary Entry, or Screenplay Format). They were asked to identify parallels, extract insights, and reflect on personal growth. After completing the task, the participants were asked to fill out the NASA Task Load Index (TLX) for cognitive load, the Post-Study System Usability Questionnaire (PSSUQ) for usability assessment, and the Creative Support Index for evaluating creative support. The participants completed a survey gathering demographic information, writing habits, and experience with AI writing tools. 

Finally, participants engaged in a retrospective interview while reviewing the snippets of the video recordings of their session. We wanted to understand the (1) affordances, in terms of effects of the application on their writing process,  (2) their embodied competencies, skills, and knowledge they drew upon during the task, and (3) any perceived rules, norms, or expectations that guided their approach. We wrapped up the session by asking about overall impressions and additional insights. The entire writing session was recorded, capturing both the participants' audio recording and the participants' screen. Each participant received \$30 for their participant in the form of an Amazon gift card.

\subsubsection{Results}
\textbf{NASA TLX.} The survey measures six subscales: Mental Demand, Physical Demand, Temporal Demand, Performance, Effort, and Frustration. Analysis of the data (N = 12) revealed varying levels of perceived workload across these dimensions. Effort ($\mu = 8.42$, $\sigma = 6.50$) and Mental Demand ($\mu = 6.17$, $\sigma = 3.43$) were the highest-rated factors, indicating that participants found the task mentally taxing and requiring substantial effort. This aligns with expectations, as the task was designed to require deep thinking and was non-trivial. Temporal Demand also had a notable score ($\mu = 6.42$, $\sigma = 6.35$), suggesting that time pressure played a significant role. \change{The Performance score ($\mu = 5.25$, $\sigma = 3.33$) indicates participants generally performed well, as lower values on this scale represent better perceived performance. Similarly, low Frustration scores ($\mu = 3.58$, $\sigma = 3.96$) suggest minimal participant frustration.} The high standard deviations, particularly for Effort and Temporal Demand, reflect significant variability in study participants' experiences, likely due to differences in their task approach or expertise characteristics. Overall, these findings suggest that \system imposed a low cognitive load on participants.


\begin{figure}[!t]
  \centering
  \includegraphics[width=0.9\columnwidth]{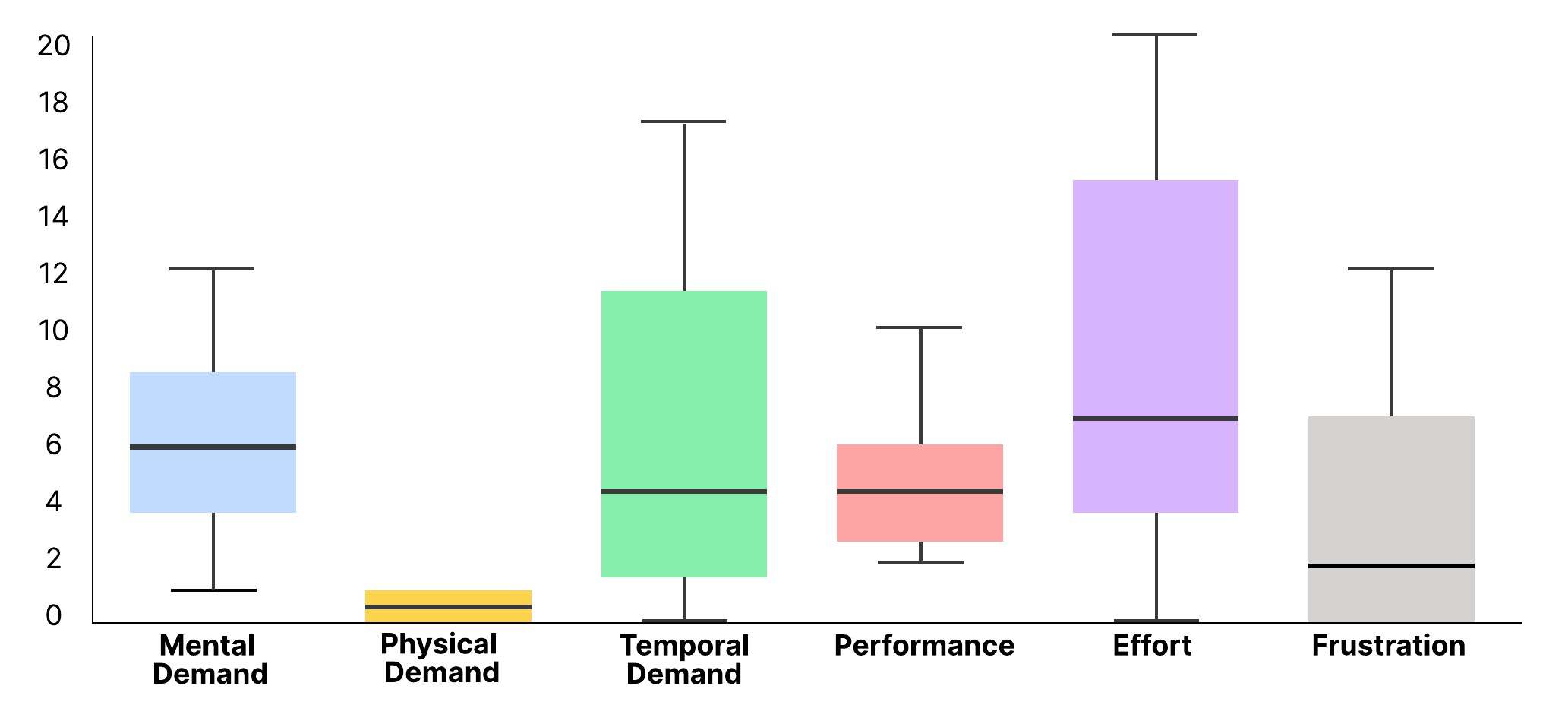}
  \caption{NASA-TLX Workload response distribution across relevant dimensions}
  \label{fig:nasatlx}
\end{figure}

\begin{figure}[!t]
  \centering
  \includegraphics[width=0.9\columnwidth]{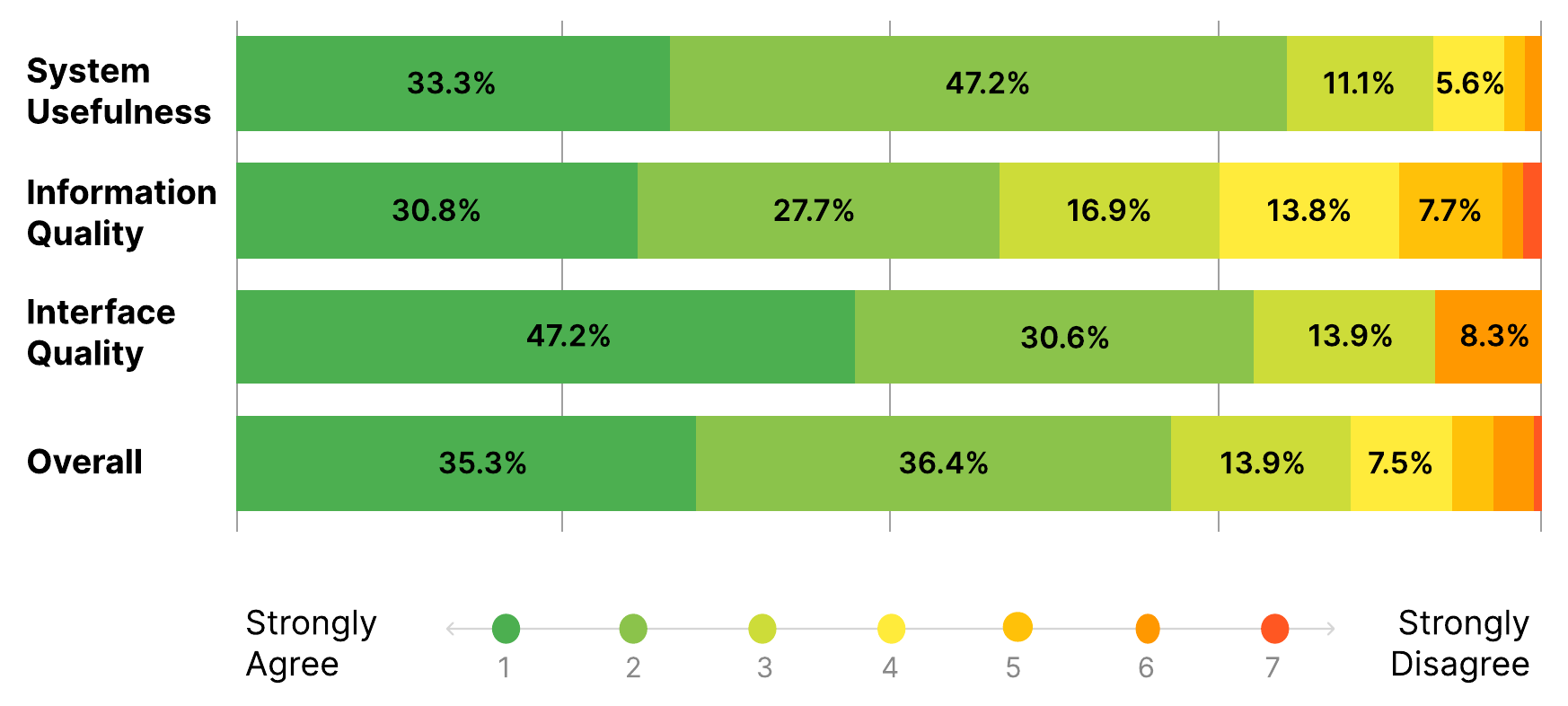}
  \caption{PSSUQ Response dimensions across System Usefulness, Information Quality, Interface Quality categories and Overall dimension}
  \label{fig:pssuq}
\end{figure}
\textbf{PSSUQ.} The Post-Study System Usability Questionnaire was used to assess the usability of the AI-assisted writing tool across three key dimensions: System Usefulness, Information Quality, and Interface Quality. Analysis of the data (N = 12) revealed generally positive usability scores, with all dimensions receiving mean ratings below the midpoint of the 7-point scale (where lower scores indicate better usability). The Overall PSSUQ score ($\mu = 2.18$, $\sigma = 1.29$) suggests that participants found the system to be reasonably usable. System Usefulness ($\mu = 1.99$, $\sigma = 1.01$) and Interface Quality ($\mu = 2.00$, $\sigma = 1.39$) were rated the highest, indicating that users found the tool functional and easy to interact with. The Information Quality dimension received a slightly lower but still positive rating ($\mu = 2.51$, $\sigma = 1.45$), suggesting room for improvement in the clarity and organization of information provided by the system.


Participants reported that they could not always tell if they had made a mistake in the interface, though this may not be a significant concern for a writing tool where many `mistakes' are subjective or stylistic choices. Within the System Usefulness dimension, items related to ease of use ($\mu = 1.67$, $\sigma = 0.89$) and efficiency ($\mu = 1.58$, $\sigma = 0.79$) received particularly positive ratings, suggesting that users found the tool intuitive and time-saving. However, the item related to system capabilities (System Usefulness: $\mu = 2.08$, $\sigma = 1.73$) showed higher variability, indicating diverse opinions on whether the system had all the expected functions and system capabilities, from the participants' perspectives.

In the Information Quality dimension, the item related to error message clarity ($\mu = 4.38$, $\sigma = 1.41$) received the lowest rating, highlighting a significant area for improvement. Interface Quality items were consistently rated positively, with low variability, suggesting a well-designed user interface. The relatively high standard deviations across most items indicate varied user experiences, possibly due to differences in expectations, prior experience with similar tools, or the specific writing tasks undertaken during the study.

\textbf{CSI.} The Creative Support Index was used to evaluate participants' experiences with \system, using a scale of 0 to 20, where lower scores indicate more positive outcomes. The analysis focused on five key dimensions: Exploration, Enjoyment, Results Worth Effort, Immersion, and Expressiveness. Data from 12 participants revealed generally positive experiences across these dimensions. Enjoyment was the highest-rated factor ($\mu = 2.67$, $\sigma = 3.50$), suggesting that users found the tool particularly enjoyable. This was closely followed by Results Worth Effort ($\mu = 3.83$, $\sigma = 3.81$), indicating that participants felt their input produced valuable results. Exploration ($\mu = 4.00$, $\sigma = 4.31$) and Expressiveness ($\mu = 4.33$, $\sigma = 3.17$) also received favorable ratings, implying that the tool effectively supported idea generation and self-expression.

\begin{figure}[!t]
  \centering
  \includegraphics[width=0.9\columnwidth]{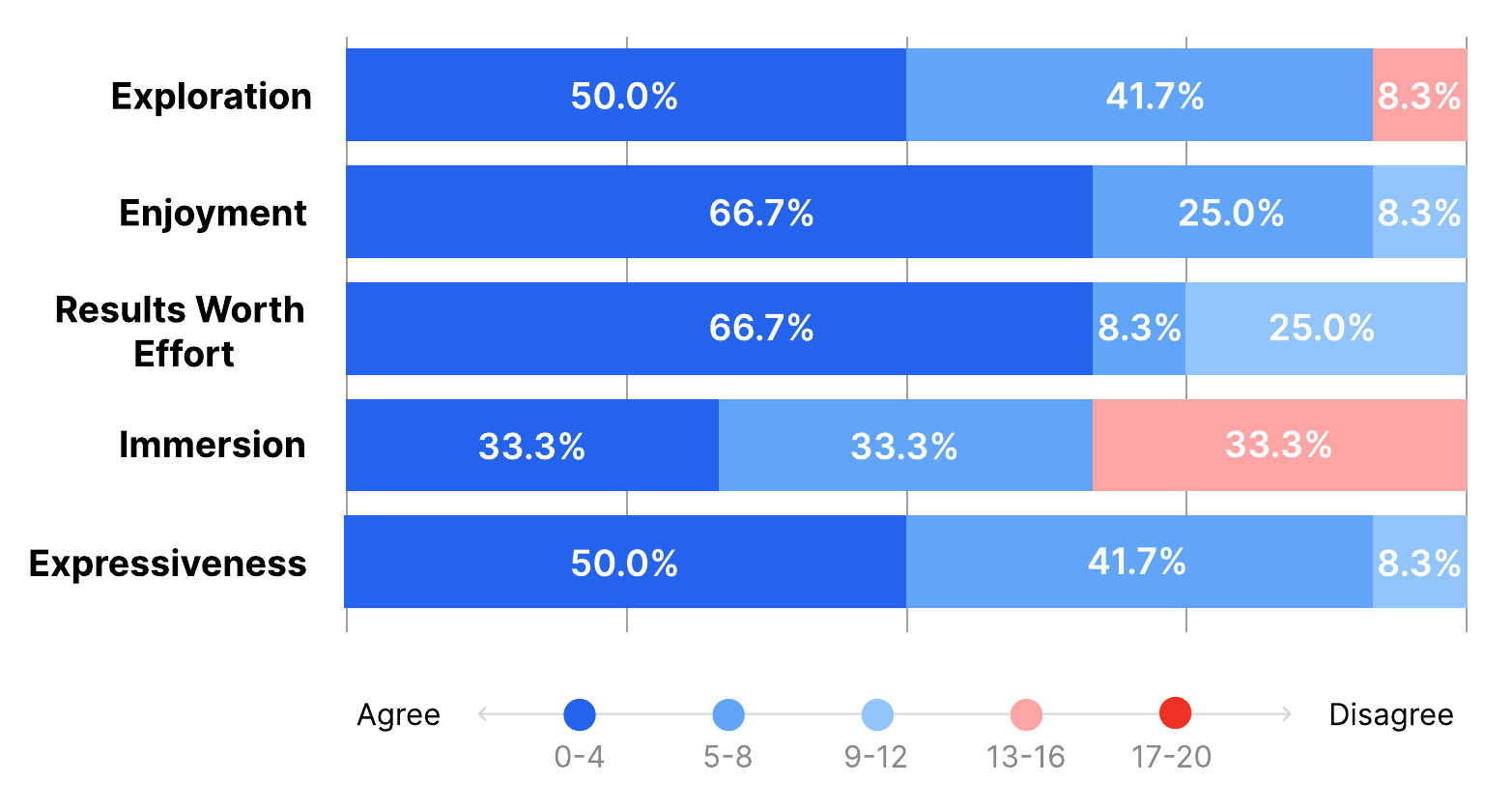}
  \caption{Creative Support Index (CSI) response distribution across relevant dimensions}
  \label{fig:csi}
\end{figure}


\begin{figure*}[t!]
    \centering
    \includegraphics[width=0.9\textwidth]{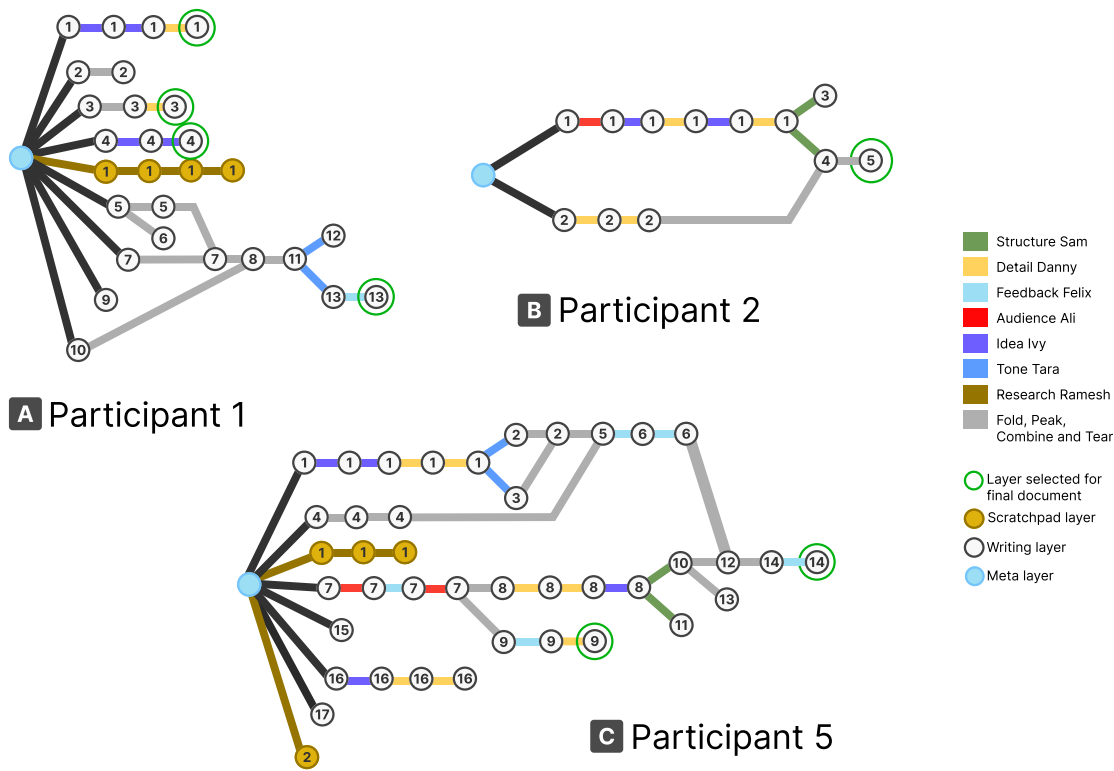}
    \caption{\change{Tree visualization of layer manipulation and LLM calls in \system. We show three diverse ways participants leveraged the interface affordances for completing the assigned task. The diverging gray lines depict `tear', the converging gray lines depict `combine', and meaning of other symbols is present in the legend}}
    \label{fig:participant-journey}
\end{figure*}

Interestingly, Immersion received the lowest rating ($\mu = 7.50$, $\sigma = 5.79$), though it still fell on the positive side of the scale. This suggests that while users found the tool engaging, there may be opportunities to enhance its ability to create a more immersive experience. The high standard deviations across all dimensions, particularly for Immersion and Exploration, reflect significant variability in user experiences, possibly due to individual differences in writing styles. The Collaboration dimension was excluded from our findings, as participants rated its weight factor as 0. 

\textbf{Spatial organization of content.} \system provides users with the ability to spatially organize their layers and associated content, a feature that was highly valued by participants for improving their focus and workflow. P9 praised the interface's immersive quality, stating: \textit{``One nice thing is that the fact that it's all in one big interface makes it less immersion-breaking than, say, opening a bunch of Google Docs tabs, where you have to make more major context switches and get distracted by other tabs.''} P9 further emphasized the advantages of a dedicated workspace: \textit{``Or you break out of the immersion of writing and being in the zone by going more into your desktop environment, where you're reminded of your work, or end up seeing social media. So it's nice to have a dedicated, isolated workspace.''}  This sentiment was shared by other participants who recognized the organizational benefits of \system over traditional text editors. P4, for instance, pointed out the limitations of conventional file systems: \textit{``The structure and organizational possibilities of this kind of thing would be huge because you end up with folder upon folder upon folder.''} P4's comment highlights \system's potential to address the organizational challenges typically encountered with standard text editors, offering a more intuitive and flexible way to manage content throughout the writing process. 

The system's spatial design supported expressive content management strategies\footnote{We thank psycholinguist George Miller, author Roy Pea's postdoctoral mentor at Rockefeller University in the 1970's for his seminal insights half a century ago in foregrounding 'the human tendency to locate information spatially', which we expressly leverage in the design of \system~\cite{millerpsychology}}. P9 emphasized how this shaped their writing process: \textit{``Being able to fold or bin my writing without permanently removing them made me much more willing to experiment. In single-page interfaces, I often feel pressured to be certain about content placement before typing.''} Users demonstrated remarkable spatial awareness of their content, organizing it in ways that enhanced accessibility. P11 noted: \textit{``Being able to [tunnel] into content from all over and get contextually-informed responses from the friends made everything feel so accessible.''} Through reduced cognitive load, users focused more on writing and experimentation rather than on content management.

\textbf{Paper metaphor for writing.} A central metaphor that \system aims to embody is the ability to manipulate and move layers much like rearranging sheets of paper on a desk. P11 captured this sentiment, stating, \textit{``It's a much more visual, like a desk with pieces of paper all over it,''} highlighting the intuitive, tactile nature of organizing content in the workspace. This visual and spatial approach offers users a more flexible and natural way to manage their writing, akin to physically handling documents in a traditional environment. It also opens up future promising research directions to develop and study a gesture- and voice-based rendition of the \system functionalities. 

\textbf{Flexibility in testing rhetorical strategies.} The affordances granted by \system, including the ability to tear, split, combine, stack, and fold layers, were clearly evident in how participants utilized the tool. These features not only supported writing tasks but also aligned with users' conceptual models of the writing process. P1 drew a parallel between the system's structure and traditional writing approaches: \textit{``Layers are like, if you're writing a paper, you need an outline and goal. Intuitively, it's like a tree which is the outline, and you work on each part of the node. In this case, you start with the intro, and you spend time collaborating with Idea Ivy. It was natural to break down [the writing process].''} This natural breakdown of the writing process was further enhanced by the system's manipulable interface, as P4 highlighted: \textit{``I really liked it. Being able to push something over there until later, and then bring it back and like smoosh it all together. That was really nice; that I liked that a lot.''} 

\change{The interface's bottom-up approach to LLM integration particularly enhanced exploratory writing. P5, whose usage journey is visualized in Fig. ~\ref{fig:participant-journey} C, observed: \textit{``Clicking a layer, moving it around and expected something to happen is so intuitive. The interface interactions felt natural and expected - exactly what I envisioned would happen.''} Participants often combined different features creatively, as P5 described: \textit{``I think order to exposition is really important. Tearing and recombining in different orders helped me rapidly see what narrative flow made sense''} P5 also wrote initially in first person, utilizing Tone Tara to shift to third person narration, demonstrating the system's flexibility in supporting various narrative styles.}

\textbf{Collaboration with Writer's Friends:} Participants demonstrated diverse and interesting approaches to leveraging the capability of the Writer's Friends in \system. These distinct personas, each representing different writing assistance features, effectively bridged the gulf of envisioning~\cite{subramonyam2024bridging}. P8 expressed a particular fondness for one such friend: \textit{``I like Danny. My Danny's a good guy.''} Similarly, P11 highlighted the value of constructive criticism: \textit{``I really like the friends, especially the feedback. I liked getting negative feedback.''} \change{Through writing in the interface and receiving feedback from the ``friends'', users frequently refined their meta layer. P2 explained: \textit{``Sometimes I felt my ideas were being misconstrued so updating the meta layer helped''} The iterative nature of getting feedback helped clarify writing goals and better understand the target readers. On the anthropomorphized LLM scaffolding, P8 remarked their indifference, ``I don't think it influenced my usage positively or negatively''} 

A thematic analysis of user-defined prompts revealed that participants accurately matched their requirements to the appropriate friend in the majority of cases. Out of the 57 times Detail Danny was used, the participants gave it prompts for detail and elaboration 77\% of the time. In the case of Idea Ivy, out of the 52 instances used across sessions 86\% of prompts were for ideation and brainstorming. In case of Tone Tara and  Structure Sam, 85\% (20) and 95\% (20) mapped to tone transformation and structuring respectively. Importantly, the Writer's Friends seemed to enhance rather than replace the creative process. P12 noted, \textit{``I honestly felt like I was still using this tool\ldots to come up with my own ideas. And I think that was good for me.''} This sentiment suggests that \system successfully balanced AI assistance with the preservation of user agency in the writing process. 

\textbf{Usability and User Interface.} The intuitive design and user-friendly interface of \system were frequently highlighted by participants, emphasizing the tool's ease of use and its ability to seamlessly integrate into the writing process. P11 expressed enthusiasm for the command interface, noting its natural feel: \textit{``I love the backslash and having the pull down menu and immediately being able to select. I liked that. I didn't even think about it\ldots I naturally did it.''} This comment showcases the effectiveness of the interface in reducing the cognitive load on users, allowing them to access features quickly and intuitively without disrupting their writing flow. The system's design also contributed to a positive emotional experience for users. P7 remarked on the sense of control and comfort provided by the interface: \textit{``I wasn't stressed at all, I felt completely in control.''}

In figure \change{~\ref{fig:participant-journey} we can see the journey through the interface taken by users with three distinct usage patterns. These visualizations give us a sense of how different writers leverage \system to test rhetorical strategies.  In the case of P2 we see that they created a total of 5 layers, two of which were alternative structures suggested by Structure Sam. In case of P1 they created 13 layers and used 4 of them in their final essay generation. Similarly, P5 created 17 layers, counting all tears, combination, and alternative suggestions layers, and used only two to generate their final essay. }

\subsection{Comparative Analysis: Between-Subjects Evaluation on Prolific}
\change{Building on the insights from the usability assessment, we designed a comparative evaluation. We wanted to understand what facet of \system supported the dynamic knowledge transformation we observed by the writers in study 1. To isolate the effects from different features, we constructed two separate interfaces in addition to \system. We conducted a between-subject study with three conditions: (1) a layered interface with in-line LLM (\system condition), (2) a writing interface with in-line LLM but no layers or spatial component (In-Line-LLM), and (3) a writing interface with a separate AI chat window for LLM interaction (Chat-LLM condition).}


\begin{figure*}[t!]
  \centering
  \includegraphics[width=0.95\textwidth]{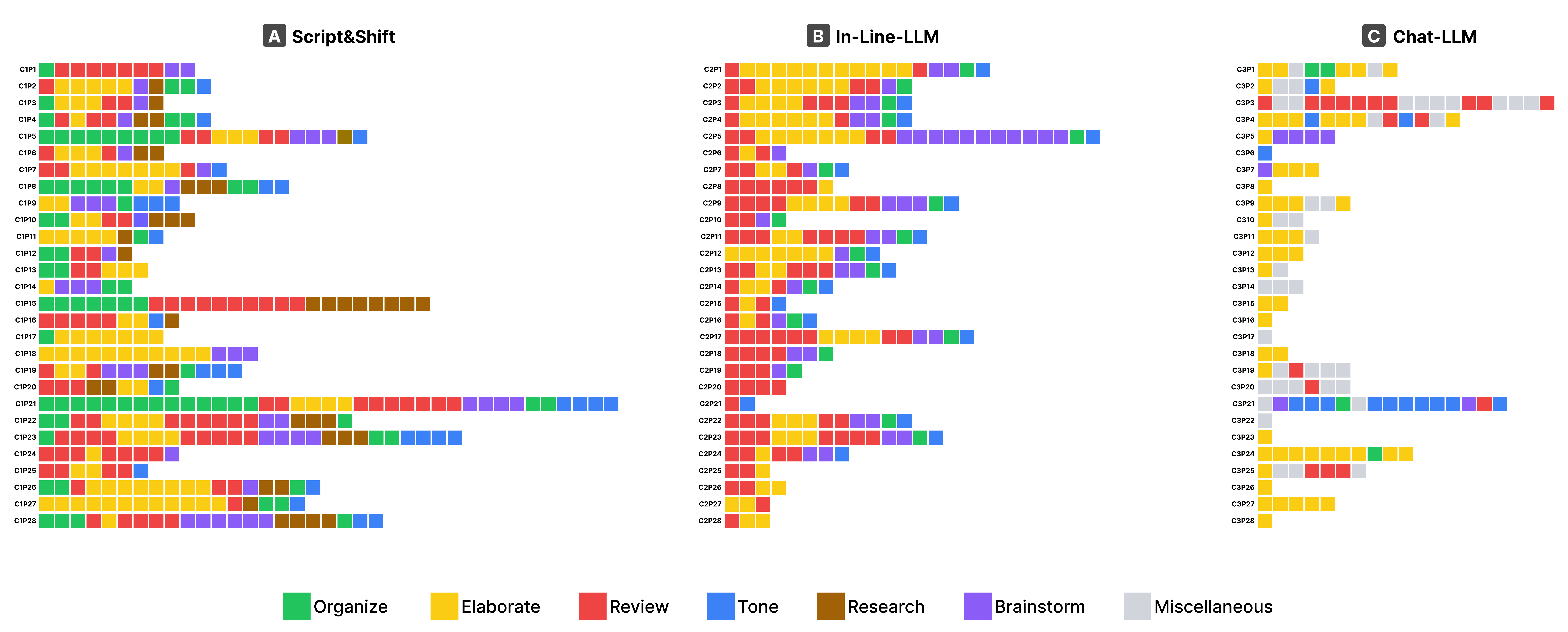}
  \caption{Visualization of LLM Interaction Across Conditions. Each square represents a different writing subprocess, with their meaning defined in the legend.}
  \label{fig:all_conditions}
\end{figure*}

\subsubsection{Design Decision For Conditions}
In order to have a fully functional In-Line-LLM application that supports the Writer's Friends without the layered paradigm, we had to make some design choices. Detail Danny, Idea Ivy, Feedback Felix and Audience Ali, have their operations constrained on the active layer so they will remain the same but in the case of Structure Sam and Tone Tara, where new layers with transformed content are generated, we had to change their behavior for the In-Line-LLM condition. Both Sam and Tara, instead of generating new layers, replaced the existing content in the editor and provided the user the option in the toolbar to reverse it to their previous content before the transformation. In the case of the Chat-LLM interface, we added a chat interface right beside the editor, to resemble having a word processor and an LLM chat interface open in split view. The rationale for creating our own interface instead of using a baseline of an existing LLM chat interface with a word processor was to have control over the data logging.

\subsubsection{Participants}
We conducted the study through Prolific and had a screening for people who wrote professionally, roles included journalist, copywriter/marketing/communications, and creative writing. We also screened for participants who spent more than 5 hours a week on Prolific, to increase chances of high-quality participation. We recruited 84 participants (F=48, M=30, NB=6). Participants reported ages between 18 and 64, the median age being in the 45-54 age group. 

\subsubsection{Study Procedure}
Each participant was randomly assigned to one of the three conditions (\system, In-Line-LLM and Chat-LLM) such that there was a distribution of 28 participants per condition.  On a scale of 1 to 5, participants reported confidence in their writing skills as follows: chat-llm ($\mu = 3.96$, $\sigma = 0.98$), in-line-llm ($\mu = 4.20$, $\sigma = 0.61$), and \system ($\mu = 4.42$, $\sigma = 0.64$) in case of \system, This study design would allow us to isolate the effect of different components and formulate a comprehensive assessment. Each participant first filled out a demographic survey, and consented to having their writing data logged into our database. The participants were each asked to watch a video tutorial for their respective condition interface. Afterwards,  they had to take a quiz about the interface. This helped ensure that the participants understood the interface well before using it. They were not allowed to proceed without getting all the answers right. Following this task they were assigned their task, which they were given 40 minutes to complete. The task required writing about an 800 words essay on a \textit{crowdworker's experiences with Large Language Models (LLMs):} They were specifically asked that it cover LLMs' impact on their work, adaptation strategies, and future outlook. They were instructed that the essay should include specific examples, data, and reflections on both positive and negative aspects of LLMs in crowd-work. The task was intentionally made non-trivial as we wanted to simulate a real writing task for the participants in a creative capacity. In order to ensure the participants did not write someplace outside of the interface before pasting it later, we tracked their activity in the interface and made persistent activity a criterion for their submission to be valid. We also manually validated the essays written to ensure they were actual essays about the topic and not something off-topic or incoherent.


\subsubsection{Results}

\change{Figure ~\ref{fig:all_conditions} shows the sequence of invocation of features across conditions. We coded the prompts issued by users into one of the following categories: (1) Review, (2) Tone Transformation, (3) Elaboration, (4) Organize, (5) Brainstorm, and (6) Research. Between the In-Line-LLM and \system condition, we combined Feedback Felix and Audience Ali into Review. We also marked prompts that did not fit into any of these categories as miscellaneous. We also do not show instances where users reissue commands when they are not content with the initial generation. Notably, analysis of prompt reissuance revealed a higher density in the Chat-LLM condition, suggesting users were frequently dissatisfied with initial responses and attempted to regenerate content. In contrast, we observed substantially lower reissuance in the In-Line-LLM and \system conditions. As we can see in the visualization, usage differences exist between the interfaces that support in-line LLM support and the chat interface. Looking at the prompts issued by the user, we observed difficulty in understanding how the LLM could support their writing. We observed prompts like, \textit{``You're not a very comprehensive model, are you?''} and\textit{  ``So, your main function is to serve as a research tool?''.} Despite providing an introduction video for all conditions, users experienced the most difficulty in the chat interface. We attribute this to challenges in formulating specific writing intentions and planning, which we believe the task-specific LLM personas were successful at bridging. Users in the In-Line-LLM and \system asked for assistance with transformation or elaboration. In the case of the chat interface, we observed users attempting to solicit complete essays based on points they specified in the prompt.}

\change{When comparing the In-Line-LLM and \system conditions, we observed greater diversity in writing strategies among \system users. A notable pattern was the tendency to seek feedback earlier in their writing process, suggesting that users aimed to align their work with audience expectations and prevent significant deviations as their drafts progressed. Conversely, the In-Line condition exhibited a more fixed usage pattern, where users typically began with review tasks, spent the bulk of their time elaborating on content, and concluded with tone transformations. Interestingly, this pattern deviated from the tutorial video shared with participants, leaving us uncertain about the underlying reasons for this behavior. Through this analysis, we conclude that \system effectively helps users bridge the envisioning and articulatory gap, enabling them to articulate their writing intentions and integrate feedback more seamlessly into their workflows. While in-line support fosters more thoughtful usage of LLM,  the spatial organization of \system propelled the dynamic knowledge transformation.}

\section{Discussion}
In this section, we reflect on the key findings from the development and evaluation of \system, while also exploring future opportunities for research into accessible media authoring tools.

\subsection{Scripting \& Shifting with LLMs}

\change{Study findings indicate that a significant advantage of \system is its ability to support non-linear and divergent writing workflows, enabling users to overcome the limitations of traditional page-based tools. Unlike conventional writing applications that enforce a rigid, sequential structure, \system employs a \textbf{layered paradigm} inspired by the concept of collage \cite{buschek2024collage} and spatial interfaces. Arguably, this approach aligns with how individuals naturally think -- separating, refining, and integrating complex ideas -- and facilitates a more fluid and exploratory writing process. At the core of \system’s layered paradigm is the ability to \textbf{script and shift} between various writing strategies seamlessly. Writers can transition effortlessly between planning, editing, structuring, and reviewing, akin to assembling a collage where different elements are combined, rearranged, and refined to create a cohesive whole. The \textbf{Writer’s Friends} further enhance this experience by providing specialized functionalities tailored to distinct phases of the writing process. Each layer functions as a cognitive sandbox, dedicated to tasks such as brainstorming with Idea Ivy, expanding ideas using Detail Danny, or reorganizing content through Structure Sam. This modularity preserves creative control and encourages users to experiment with different rhetorical strategies and content structures without the constraints of a linear workflow. Further, it provides writers with affordances for iterative and divergent writing process, essential for creative work \cite{sawyer2021iterative}}

\change{A key feature that participants found compelling was \textbf{non-destructive editing}. \system allows layers to be folded, discarded, or rearranged without permanently deleting any content, providing writers with the freedom to experiment without the fear of losing their work. This capability is crucial for exploring various tones, audience alignments, and structural organizations. The \textbf{spatial organization} of content within \system also plays a key role in supporting creative workflows. By visualizing content in a layered, spatially oriented manner, writers can perform cross-content comparisons, experiment with folding or reorganizing layers, and engage in ``tear, combine, and peek'' sequences of transformations. For instance, participant P3 described an iterative process of tearing apart a layer, reordering its fragments, and recombining them to evaluate the impact on the essay’s overall flow. This spatial manipulation mirrors the collage-like assembly of ideas, allowing writers to focus on individual aspects of their work while simultaneously exploring broader structural or stylistic variations. Participants appreciated the ease with which they could swap out layers and assess the influence of their presence or absence on the narrative, facilitating a deeper understanding of how different elements interact within their writing.}

\change{In summary, \systems integration of the layered paradigm and collage-inspired affordances creates a robust framework for non-linear and divergent writing workflows. By supporting non-linear, spatially-oriented content management, \system reduces cognitive friction to 'writing ballistics'~\cite{peakurland} and fosters an environment conducive to both exploratory and goal-directed writing. This alignment between user mental models and system functionality underscores the importance of designing writing interfaces that prioritize spatial organization and direct manipulation of text.}

\subsection{Broader Utility and Non-Linear Workflows}
While \system has demonstrated its value in supporting creative writing workflows, its broader utility extends beyond individual use cases. One particularly promising area is in education, where the layered interface and LLM-assisted support could transform how writing is taught and practiced. \systems ability to scaffold \cite {pea2018social} the writing process, from brainstorming and outlining to drafting and revising, makes it an ideal tool for educators seeking to help students develop their writing skills in a more exploratory, iterative manner aligned with the cultural practices of expert writers.

Beyond education, \system holds potential for other forms of content creation, particularly where non-linear thinking is key. In areas like multimedia storytelling, screenwriting, or journalism, \system’s layered interface could be used to storyboard, organize interviews, or manage multimedia elements. Alternative designs building on the \system metaphor could be developed for other cognitive design activities such as songwriting, screenplay writing, and movie scripting. The ability to work on different parts of a project in parallel, without being tied to a linear structure, could greatly enhance workflows in these disparate fields. For instance, a journalist might use \system to organize interviews, drafts, and research into layers, while screenwriters could draft scenes in a non-linear order, rearranging them to find the optimal narrative structure.

The findings from this study demonstrate that \system effectively supports non-linear, divergent writing processes through its layered, LLM-enhanced workspace. A key strength of LLMs, as evidenced in our study, is their ability to generate \textit{diverse and unexpected alternatives}, allowing users to experiment with multiple approaches in real-time. This makes \systems layered paradigm potentially useful in other domains such as policy development, screenwriting, and even generative programming (e.g.,~\cite{angert2023spellburst}). For instance, in policy writing, \system’s ability to help users generate and compare alternative argument structures, draft revisions, and outline complex documents in layers offers significant advantages. Screenwriters could leverage LLMs to explore multiple versions of dialogue, scene settings, or character interactions, helping them iterate on scripts faster while maintaining creative control over the process. They could use template features in \system to experiment with different narrative arcs while developing their character and scene structures. By enabling users to shift between drafting, editing, and comparing different versions of their work, \system provides a versatile platform that supports not only creativity but also structured, iterative workflows across diverse fields.

\subsection{Generalizability}
\change{As discussed above, by decoupling content development and structural organization into modular layers, \system can accommodate diverse workflows, from policy analysis to creative media production. This \textit{separation of concerns} allows the architecture to support non-linear and iterative processes, which are crucial for different genres, purposes, and user preferences. By structuring content as discrete, manipulable layers, the architecture facilitates intuitive spatial reasoning, enabling users to experiment with content and structure without compromising the coherence of their overall work enterprise. Interfaces based on this architecture could move beyond traditional document-centric models to offer dynamic, zoomable workspaces where users interact with content at multiple levels of abstraction, seamlessly shifting between granular edits and high-level structural organization.}

\change{Additionally, the architecture's extensibility through ``Writer's Friends'' makes it adaptable to varied contexts and rhetorical goals. For example, AI agents could operate within specific layers, offering suggestions tailored to the scope of the layer (e.g., tone adjustments for a narrative layer or argument alignment for a reasoning layer). At a higher level, cross-layer AI could evaluate coherence, transitions, or overarching goals, supporting interfaces that bridge the micro and macro aspects of complex projects. Finally, the architecture could also support interfaces optimized for collaborative environments, where individual contributors interact with distinct layers representing their areas of focus, supporting both autonomy and coherence.}

\subsection{Limitations and Future Work}
While \system demonstrates clear advantages in supporting flexible writing workflows, some limitations emerged during the study. One challenge lies in the initial learning curve associated with the layered interface. Some participants, particularly those with less experience using non-traditional writing tools, found the spatial organization and multi-layered features overwhelming initially. Future iterations could include enhanced onboarding and tutorials to progressively ease users into mastering the system’s capabilities. A second limitation relates to the variability in how participants used Writer’s Friends. While most found them helpful, some struggled to select the appropriate friend for their task, or felt that the generated content did not always align with their specific writing goals. This finding suggests room for improvement in tailoring the system’s AI outputs to a wider variety of writing styles and preferences. Finally, although \system is designed to preserve user voice and reduce cognitive load through in-line AI assistance, there remains a risk of LLM bias in the generated content. For instance, a few participants were skeptical of ``Research Ramesh,'' citing distrust in LLMs due to their tendency to fabricate information. To address this, the system could require users to provide context documents, restricting LLM responses to verified excerpts. While the current implementation relies on commercial LLM safety measures, integrating chain-of-thought reasoning could further reduce hallucination~\cite{ji2024chain} and bias by ensuring more transparent and grounded outputs. While the separation of user input and AI-generated content helps maintain transparency, future versions could explore ways to provide more control over LLM-generated outputs and increase the transparency of how content is produced.

\section{Conclusion}

\system introduces a novel \textit{layered interface paradigm} designed to overcome the limitations of traditional writing tools. By allowing writers to manage content development and structural organization through a flexible, zoomable workspace, \system supports seamless interaction between micro and macro writing tasks. Its layered design enables non-linear editing, divergent thinking, and the integration of LLM-generated content without disrupting the writer's workflow. Features like stacking, folding, and the use of Writer's Friends help reduce the cognitive load and improve the fluidity of writing to create knowledge-in-action \cite {bamberger1983learning}. Our user study demonstrated that participants appreciated these capabilities, which facilitated their more exploratory and efficient writing process than what they usually experienced. This highlights \systems potential to enhance creativity and streamline complex writing tasks.

\begin{acks}
  We are grateful to the reviewers and our study participants for their time and helpful feedback. We also thank Tyler Angert, Jasmine Shih, Sarah Levine, and Alec Helbling for their feedback on the system design and help with figures.  This work is supported through the AI Research Institutes program by the National Science Foundation and the Institute of Education Sciences, U.S. Department of Education through Award $\#2229873$ - National AI Institute for Exceptional Education.
\end{acks}

\bibliographystyle{ACM-Reference-Format}
\bibliography{99_refs}

\end{document}